\documentclass[%
 reprint,
superscriptaddress,
 amsmath,amssymb,
 aps,
 pra,
]{revtex4-2}

\usepackage{dcolumn}
\usepackage[colorlinks=true, allcolors=blue]{hyperref}
\usepackage{braket}
\usepackage{bm}
\usepackage{amsfonts}
\usepackage{mathrsfs}
\usepackage{mathtools}
\usepackage{amsthm}
\usepackage{graphicx}
\usepackage{color}
\usepackage{array}\usepackage[makeroom]{cancel}
\usepackage{changes}
\usepackage{qcircuit}
\usepackage{bbm}
\usepackage{comment}
\usepackage{multirow}
\usepackage{cellspace} 
\usepackage{tikz}
\usepackage{mdframed}
\usepackage{algorithm}
\usepackage{algpseudocode}
\usetikzlibrary{positioning, shapes, arrows, fit}
\usetikzlibrary{arrows.meta}
\usepackage{orcidlink}
\usepackage{leftindex}
\usepackage{empheq}
\usepackage{booktabs}
\usepackage{afterpage}
\usepackage{xcolor}
\usepackage{booktabs, makecell}
\usepackage[capitalize]{cleveref}

\crefname{thm}{Theorem}{Theorems}
\crefname{dfn}{Definition}{Definitions}
\crefname{rmk}{Remark}{Remarks}
\crefname{lem}{Lemma}{Lemmas}
\crefname{cor}{Corollary}{Corollaries}

\theoremstyle{plain}

\theoremstyle{remark}

\newcommand{\SOC}{\text{SOC}}

\newcommand{\ISC}{\text{ISC}}
\newcommand{\eff}{\text{eff}}

\begin{document}


\title{Quantum algorithms to identify optically detected magnetic resonance-active defects for quantum sensing applications}

\author{Pablo A. M. Casares \orcidlink{0000-0001-5500-9115}}
\affiliation{Xanadu, Toronto, ON, M5G 2C8, Canada}

\author{Yanbing Zhou 
\orcidlink{0000-0001-7673-771X}}
\affiliation{Xanadu, Toronto, ON, M5G 2C8, Canada}

\author{Utkarsh Azad
\orcidlink{0000-0001-7020-0305}}
\affiliation{Xanadu, Toronto, ON, M5G 2C8, Canada}

\author{Stepan Fomichev \orcidlink{0000-0002-1622-9382}}
\affiliation{Xanadu, Toronto, ON, M5G 2C8, Canada}

\author{Jack S. Baker \orcidlink{0000-0001-6635-1397}}
\affiliation{Xanadu, Toronto, ON, M5G 2C8, Canada}

\author{Chen Ling \orcidlink{0000-0002-8667-9792}}
\affiliation{Toyota Research Institute of North America, Ann Arbor, MI, 48105, USA}

\author{Debasish Banerjee}
\affiliation{Toyota Research Institute of North America, Ann Arbor, MI, 48105, USA}

\author{Alain Delgado \orcidlink{0000-0003-4225-6904}}
\affiliation{Xanadu, Toronto, ON, M5G 2C8, Canada}

\author{Juan Miguel Arrazola \orcidlink{0000-0002-0619-9650}}
\affiliation{Xanadu, Toronto, ON, M5G 2C8, Canada}

\begin{abstract}
Spin defects in two-dimensional materials are a promising platform for quantum sensing. 
Simulating the defect's optical response and optically detected magnetic resonance (ODMR) contrast is key to identifying suitable candidates. However, existing simulation methods are typically unable to supply the required accuracy.
Here, we propose two quantum algorithms to detect an imbalance in the triplet-to-singlet intersystem crossing (ISC) rates between excited states with the same and different spin projections---a necessary condition for nonzero ODMR response.
The lowest-cost approach evaluates whether the evolution of an $S=0$ state under the spin-orbit coupling induces ISC to $S=1$, and also whether there is an imbalance in its intensity depending on the final state spin projection.
The second approach works by comparing the emission spectrum of a spin defect with and without the spin-orbit coupling operator, inferring ISC intensity for different spin transition channels from spectrum intensity changes. 
Additionally, we present an improved scheme to evaluate the defect's optical response, building upon previous work.
We study these quantum algorithms in the context of the negatively charged boron vacancy in hexagonal boron nitride. We generate an embedded active space of 18 spatial orbitals using quantum defect embedding theory (QDET) and show that the ISC rate imbalance can be detected with as few as 105 logical qubits and $4.41 \times 10^8$ Toffoli gates.
By avoiding direct and costly rate calculations, our methods enable faster screening of candidate defects for ODMR activity, advancing the prospect of using quantum simulations to aid the development of high-performing sensing devices.
\end{abstract}


\maketitle


\section{Introduction}

Recent advances in creating point defects in two-dimensional materials have sparked efforts to develop ultra-sensitive sensors capable of detecting minute magnetic and electric fields, as well as atomic-scale strains~\cite{fang2024quantum, esmaeili2024evolution, scholten2024multispecies, liu2022spinhBN}. Unlocking these applications requires identifying defects with high-spin states, sharp photoluminescence (PL) lines, and weak phonon sidebands~\cite{dreyer2018first}. Crucially, suitable defects must enable optical spin initialization and reliable readout via spin-dependent decay processes across a wide range of values for the sensed quantity~\cite{Gali2023-strain-E, stern2022}.

Optically detected magnetic resonance (ODMR) spectroscopy is the primary experimental technique for probing spin polarization in quantum defects~\cite{gottscholl2021room}. It uses a laser to pump optical excitations in the defect and microwave radiation to induce magnetic transitions between the spin sublevels. The change in PL intensity with respect to the microwave frequency is recorded in the ODMR spectrum. When a valley or peak is observed, the defect is considered ODMR-active, and the resonant frequency is used to calibrate the quantum sensor~\cite{stern2022}.

Identifying ODMR-active defects has proven to be a challenge~\cite{frey2020machine, wolfowicz2021quantum}. It is not feasible to experimentally investigate all possible defects in different host materials~\cite{bassett2019quantum, ping2021computational}. This limitation naturally motivates a simulation-guided approach to the search.
To be useful, simulations must accurately predict radiative and intersystem crossing (ISC) rates between spin sublevels, which are key to predicting ODMR-activity in spin defects~\cite{ping-simulating-odmr}.
Achieving high accuracy in simulating radiative rates, and ISC rates driven by the spin-orbit coupling (SOC) interaction, requires access to the many-electron correlated states of the defect~\cite{ping-simulating-odmr, dreyer2018first, bassett2019quantum}.
However, most current computational simulations of new defects use density functional theory (DFT) methods~\cite{qpod2022quantum, gao2021radiative}, which cannot accurately describe multi-reference states.
At the same time, the application of more advanced post-Hartree-Fock wave function methods~\cite{jensen2017introduction} that are able to treat electronic correlation is computationally prohibitive, on account of the large size required for the defect-material supercell. To make the application of these methods tractable, researchers typically employ complete active space (CAS) based approaches, which neglect important screening effects due to the host material~\cite{barcza2021dmrg, ivady2020ab}. Time-dependent DFT has also been used to calculate ISC rates~\cite{de2019predicting}. However, in this approach the Hilbert space used to represent the correlated states is restricted to singly-excited configurations, and its accuracy is limited by both the choice of the approximate density functional and the adiabatic approximation~\cite{lacombe2020developing}.

\begin{figure*}[t]
    \centering
    \includegraphics[width=0.8\textwidth]{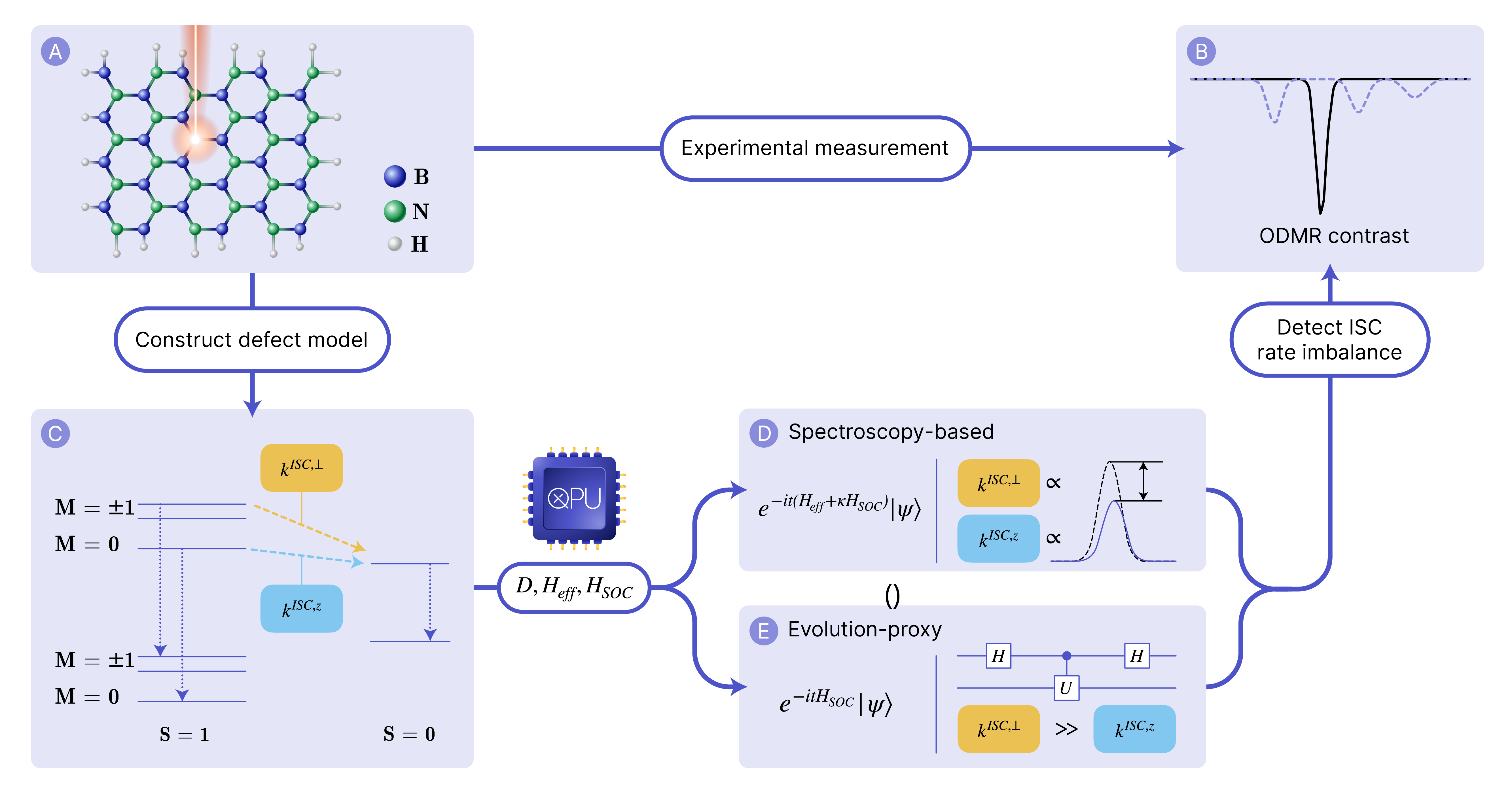} 
    \caption{Schematic representation of the workflow used to identify ODMR-active defects. (a) A laser pumps optical excitations in the defect while a microwave field induces transitions between the spin sublevels with $M=0$ and $M=\pm 1$. (b) A schematic representation of the observed ODMR contrast measured experimentally, where the dashed and solid curves correspond to different applied magnetic field strengths (see \cref{eq:odmr_contrast} for the definition of this quantity) (c) The defect's low-lying energy spectrum, coming from a QDET effective Hamiltonian, showing both radiative transitions and ISC transitions mediated by SOC, with non-axial ($k^{\text{ISC},\perp}$) and axial ($k^{\text{ISC},z}$) rates highlighted. (d) The spectroscopy-based algorithm detects ODMR activity by detecting drops in spectral intensity indicative of nonzero ISC rates, then comparing the deduced ISC rates between spin-preserving and spin-flipping channels and inferring ODMR activity from an imbalance. (e) The evolution-proxy algorithm computes ISC rate proxies by time evolving under the SOC operator for the different spin channels, and compares the relative strengths of those proxies to deduce an ISC rate imbalance and thus ODMR-activity. 
    }
    \label{fig:hero}
\end{figure*}

Quantum computing has the potential to overcome the limitations faced by classical methods for simulating spin defects. Quantum simulation of the full supercell is very costly. However, quantum defect embedding theory (QDET) can be leveraged to significantly reduce the size of the defect Hamiltonian while still accounting for dielectric screening effects from the surrounding material~\cite{sheng2022green}. Baker \textit{et al.} \cite{baker2024simulating} combined QDET with quantum phase estimation (QPE) to sample the optically active excited states of a negative boron vacancy in hexagonal boron nitride. However, estimated resource requirements revealed the need for over 1,000 logical qubits and T-gate requirements on the order of $10^{9}$ just to compute the largest dipole transition amplitude. This poses a challenge for the practical application of the algorithm.

This work introduces a more cost-effective method for determining the optical response of spin defects and identifying defects that are ODMR-active. For the optical response, we propose a modification to our previous scheme described in Ref.~\cite{baker2024simulating}. Our updated approach replaces qubitization and quantum phase estimation with a more qubit-efficient combination of Trotter formulas and the time-domain algorithm from~\cite{fomichev2024simulating,fomichev2025affordable}. 

To detect ODMR activity, we propose two quantum algorithms that compare the relative strengths of different intersystem crossing (ISC) channels, a key factor in identifying ODMR-active defects~\cite{ping-simulating-odmr}.
We refer to the first of these as the \textit{evolution-proxy} method: instead of directly computing the ISC rates, we time-evolve the system under the SOC operator and track how fast it modifies the spin of the system, as a function of the initial state spin projection. This process generates proxy quantities that are proportional to ISC rates at short times. Detecting an imbalance between these proxies allows for the inference of ODMR activity.
The second approach is an extension of a recently developed method~\cite{fomichev2024simulating} to probe the optical response of quantum systems, which we term the \textit{spectroscopy-based} approach. This method is based on the observation that including the SOC operator into the system Hamiltonian will modify the simulated emission spectrum, provided that ISC is occurring. Comparing the emission spectra with and without SOC for different initial states, we can link changes in the intensity of emission peaks to the presence of nonzero ISC rates. Furthermore, unlike the evolution-proxy method, the spectroscopy-based approach can identify specific excited states involved in ISC. This identification is based on observing which peaks in the spectrum lose intensity.
Having introduced these algorithms, we apply them to a prototypical quantum sensing system, the negatively charged boron vacancy in hexagonal boron nitride, to validate their correctness and evaluate their cost. We confirmed the algorithms' ability to correctly predict an ISC rate imbalance and accurately recover the emission spectrum by running them on the PennyLane \texttt{lightning} quantum simulator backend~\cite{bergholm2018pennylane,asadi2024}. Through constant-factor resource estimation, we find that we can predict ODMR activity in an active space of 18 spatial orbitals with as few as 105 logical qubits and maximum circuit size of $4.41 \times 10^8$ Toffoli gates (see \cref{tab:resources_algorithms}).

\begin{table*}[t]
\centering
\setlength{\tabcolsep}{9pt} 
\renewcommand{\arraystretch}{1.2}
    \begin{tabular}{c c c c c c c c}
        \multicolumn{2}{c}{} & \multicolumn{2}{c}{Evolution-proxy (per circuit)} & \multicolumn{2}{ c }{Spectroscopy (per spectrum)} & \multicolumn{2}{c}{Spectroscopy (costliest circuit)} \\ 
    \cline{3-8}
        $N$ & Qubits & Toffoli gates & Active volume  & Toffoli gates & Active volume & Toffoli gates & Active volume \\
        \hline
        \hline
        14 & $97$ & $2.07\times 10^{8}$ & $1.13\times 10^{10}$ & $5.05\times 10^{12}$ & 
        $2.77\times 10^{14}$ & $1.03\times 10^{9}$ & $5.64\times 10^{10}$\\
        16 & $101$ & $3.09\times 10^{8}$ &  $1.69\times 10^{10}$& $7.59\times10^{12}$ & $4.17\times 10^{14}$ & $1.54\times 10^{9}$ & $8.47\times 10^{10}$\\
        18 & $105$  & $4.41\times10^{8} $ & $2.41\times10^{10}$ & $1.09\times 10^{13}$ & $5.96\times 10^{14}$ & $2.21\times 10^{9}$ & $1.21\times 10^{11}$\\
        36 & $141$ & $3.56\times 10^{9}$ & $1.94\times 10^{11}$ & $8.86\times 10^{13}$ & $4.87\times 10^{15}$ & $1.80\times 10^{10}$ &$9.89\times 10^{11}$ \\
        \hline
    \end{tabular}
    \caption{Resource estimates for obtaining the ISC rate imbalance of $V_B^-$ in active spaces of $N$ spatial orbitals, using the evolution-proxy and spectroscopy-based algorithms. The optical response algorithm is the same as the spectroscopy-based algorithm. The cost of the former is slightly lower because the one-body fragment becomes spin-symmetry preserving as it lacks $H_{\text{SOC}}$. The active volume is computed with Table I from Ref.~\cite{litinski2022active}. The qubits required include those dedicated to state preparation using the Sum of Slaters technique ~\cite{fomichev2024initial}, and $D = 10^4$ Slaters. The spectroscopy algorithm requires 5 qubits less than indicated because it needs to prepare a state with $D$ Slaters, instead of the $2D$ Slaters needed in the evolution-proxy algorithm and indicated in the table, see~\cref{fig:modified_Hadamard}. The costs for the evolution-proxy algorithm assume that the dipole excites over an energy range such that $\lambda = 5$ Ha, the transition range is $\Delta = 0.05$ Ha, and the Trotter error $\epsilon_{\text{HS}} \leq \Delta$ (see~\cref{fig:trotter_error} and~\cref{eq:cost_evolution_proxy}).}
\label{tab:resources_algorithms}
\end{table*}

The rest of this manuscript is structured as follows.~\cref{ssec:odmr_basics} describes the basics of computing the ODMR contrast, and defines the optical emission and ISC rates, which are the main observables for the quantum algorithm. The defect Hamiltonian, and the dipole and SOC operators required for the quantum simulations, are defined in~\cref{ssec:obs}, constructed using QDET.~\cref{sec:algos} describes the evolution-proxy and spectroscopy-based quantum algorithms for detecting ODMR activity, as well as an improved approach to obtaining the defect's optical response. The resources needed to run the quantum algorithms are analyzed in~\cref{sec:resource_estimation}. The algorithms are applied to simulate a negatively charged boron vacancy in an hexagonal boron nitride cluster in~\cref{sec:validation}. Finally,~\cref{sec:conclusions} summarizes the main conclusions.

\section{Theory}
\label{sec:theory}

\subsection{ODMR contrast}
\label{ssec:odmr_basics}
\begin{figure}
    \centering
    \begin{tikzpicture}

  \node[font = \large] at (1.5,-0.3) {$S=1$};
  \node[font = \large] at (5,-0.3) {$S=0$};

\draw[thick] (0.5, 0.1+0) -- (2.5, 0.1+0);
\draw[thick] (0.5, 0.4+0) -- (2.5, 0.4+0);
\draw[thick] (0.5, 0.5+0) -- (2.5, 0.5+0);

\draw[thick] (0.5, 3.+0) -- (2.5, 3.+0);
\draw[thick] (0.5, 3.5+0) -- (2.5, 3.5+0);
\draw[thick] (0.5, 3.6+0) -- (2.5, 3.6+0);

  \draw[thick] (4, 1.) -- (6, 1.);
    \draw[thick] (4, 2.8) -- (6, 2.8);

\path (1.25, 0.44) coordinate (A); 
\path (1.25, 3.5) coordinate (B); 
\draw[thick, solid, <-, color = black] (A) -- (B) node[midway, left,  color=black] {}; 

  \path (1.75, 0.1) coordinate (A); 
  \path (1.75, 3.0) coordinate (B); 
\draw[thick, solid, <-, color = black] (A) -- (B) node[midway, right,  color=black] {$k^r$}; 

  \path (5, 1) coordinate (A); 
  \path (5, 2.8) coordinate (B); 
  \draw[thick, solid, <-, color = black] (A) -- (B) node[midway, right,  color=black] {$k^r$}; 

    \path (2.5, 3.5) coordinate (A); 
  \path (5, 2.8) coordinate (B); 
  \draw[thick, dashed, ->] (A) -- (B) node[midway, above, color=black] {$k^{\text{ISC},\perp}$}; 

  \path (2.5, 3.) coordinate (A); 
  \path (4, 2.8) coordinate (B); 
  \draw[thick, dashed, ->] (A) -- (B) node[midway, below, color=black] {$k^{\text{ISC},z}$}; 

  \draw[-{Triangle[length=3pt, width=3pt]}, thick, gray] (1.55, 3.25) arc (180:0:0.2) node[midway, shift={(0.7cm,-0.2cm)}, color=black] {$\omega_{\text{MW}}$};
  \draw[-{Triangle[length=3pt, width=3pt]}, thick, gray] (1.95, 3.25) arc (0:-180:0.2);

   \node[left] at (0.4, 0.45) {$M = \pm 1$};
   \node[left] at (0.4, 0.1) {$M = 0$};

   \node[left] at (0.4, 3.55) {$M = \pm 1$};
   \node[left] at (0.4, 3.) {$M = 0$};

\end{tikzpicture}
     \caption{Sketch of the lowest-lying energy spectrum of a quantum defect with a triplet ground state with spin quantum numbers $S, M$. The spin-conserving emission rate, and the axial and non-axial intersystem crossing rates are respectively denoted by $k^r$, $k^{\text{ISC},z}$ and   $k^{\text{ISC},\perp}$. Rabi frequency of the microwave radiation inducing magnetic transitions between the spin sublevels of the excited states is denoted by $\omega_\text{MW}$.}
     \label{fig:levels_rates}
\end{figure}
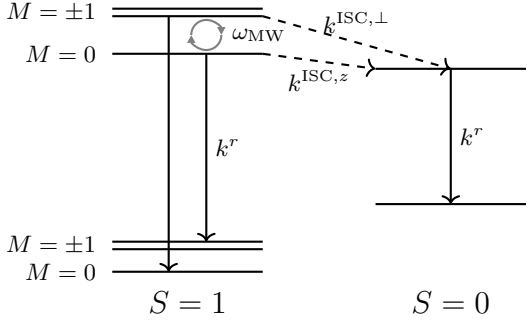

The ODMR contrast $C$ is given by~\cite{ping-simulating-odmr}
\begin{equation}
C(\omega_{\text{MW}}) = 1- \frac{\bar I(\omega_\text{MW})}{\bar I(\omega_\text{MW}=0)},
\label{eq:odmr_contrast}
\end{equation}
where $\bar I(\omega_\text{MW})$ and $\bar I(\omega_\text{MW}=0)$ are respectively the average intensity of the photoluminescence (PL) with and without the microwave (MW) field with Rabi frequency $\omega_\text{MW}$. The PL intensity is given by~\cite{ping-simulating-odmr}
\begin{equation}
\bar I(\omega_{\text{MW}}) = \sum_{i \in \text{ES}} \sum_{j \in \text{GS}} k_{ij}^r(\omega_{\text{MW}}) \bar n_i(\omega_{\text{MW}}),
\label{eq:I_PL}
\end{equation}
where $i$ and $j$ are indices running over the spin sublevels of the excited (ES) and ground states (GS), respectively, and $k_{ij}^r$ denotes the radiative transition rates. The average populations $\bar n_i$ of the excited states $\ket{E_i}$ are obtained by solving classical rate  equations~\cite{tetienne2012magnetic}
\begin{equation}
\frac{dn_i}{dt} = \sum_l (k_{li}n_l - k_{il}n_i)
\label{eq:rate_eqs}
\end{equation}
at the steady state ($dn_i/dt = 0$). In~\cref{eq:rate_eqs} the index $l$ runs over the considered defect states. The rate $k_{li}$ is determined by the amplitude $|\bra{E_l} \hat{O}\ket{E_i}|^2$, where $\hat{O}$ is an operator representing either the electron-photon coupling for radiative transitions, the electron-phonon coupling for internal conversion, or the spin-orbit coupling for intersystem crossing~\cite{ping-simulating-odmr}. Here, we focus on the optical emission $k^r$ and the nonradiative intersystem crossing (ISC) $k^\text{ISC}$ rates between the defect states, as sketched in~\cref{fig:levels_rates}. Internal conversion processes mediated by the electron-phonon interaction are not considered, as their rates are significantly smaller~\cite{ping-simulating-odmr}. 

The emission rate of spin-conserving transitions between the excited and ground states is given by~\cite{baker2024simulating}
\begin{equation}
k_{ij}^r = \frac{4}{3} [\alpha (E_i-E_j)]^3~|{\bra{E_i}} \bm {D} \ket{E_j}|^2,
\label{eq:rad_rate}
\end{equation}
where $\alpha$ is the fine-structure constant, $E_i$ denotes the energy of the state $\ket{E_i}$, and $\bm{D}$ is the electric dipole operator defined in \cref{ssec:obs}. The ISC rate is evaluated as~\cite{ping-simulating-odmr, ping2021computational}
\begin{equation}
k_{fi}^\text{ISC} = 2 \pi g |\bra{E_f} H_\text{SOC} \ket{E_i}|^2 X_{if}(T),
\label{eq:ISC_rate}
\end{equation}
where $\ket{E_i}$ and $\ket{E_f}$ are respectively the initial and final states, $g$ is the degeneracy factor of the final state~\cite{alkauskas2014first}, $H_\text{SOC}$ is the spin-orbit coupling (SOC) operator (see~\cref{ssec:obs}), and $X_{if}(T)$ is the temperature-dependent phonon contribution determined by the overlap of the initial and final vibrational states~\cite{ping-simulating-odmr}. 

From the definition of the ODMR contrast in \cref{eq:odmr_contrast}, we see that it is only nonzero when the MW field significantly modifies the PL intensity. Two conditions need to be satisfied for this to occur (see \cref{fig:levels_rates}). The first is that the MW frequency $\omega_{\text{MW}}$ needs to resonate with the energy splitting of the spin sublevels, so it can start populating excited states with nonzero spin projection. However, this in itself does not guarantee a nonzero ODMR contrast. If the ISC rates from states with different spin projection are the same, the net ISC rate is unaffected by the application of the MW field, meaning that PL is unchanged and thus the contrast will still be zero. 

The second necessary condition is thus that the excited states in the spin projection manifolds have unequal intersystem crossing rates. These are marked as $k^{\text{ISC}, z}$ for the \textit{axial} ISC rate from the $\ket{S=1,M=0}$ state to the $\ket{S=0}$ state that preserves the spin projection; and $k^{\text{ISC}, \perp}$ for the non-axial ISC rate from $\ket{S=1,M=\pm1}$ to $\ket{S=0}$ that changes the spin projection. In the case of unequal rates, the application of the MW field at the right frequency will drive some of the state population to the $M = \pm 1$ state manifold. From this manifold, transitions to the singlet state will occur at a different rate than from the $M = 0$ manifold, resulting in a noted change to the overall ISC rate and thus the PL, hence producing a nonzero ODMR contrast. This was demonstrated by the authors in Ref.~\cite{ping-simulating-odmr} who performed numerical simulations of the ODMR spectrum for the NV center in diamond. Crucially, they observed that the difference between the axial and non-axial ISC rates determines the intensity of the ODMR contrast.
For this reason, in this work, we focus on developing quantum algorithms to \textit{probe the ratio} of the axial and non-axial ISC rates.

\subsection{Observables}
\label{ssec:obs}
Simulating the optical properties of the defect as well as the intersystem crossing requires building the electronic Hamiltonian, the electric dipole, and spin-orbit coupling operators. We use quantum defect embedding theory (QDET)~\cite{sheng2022green} to construct the effective defect Hamiltonian. The main steps to build the QDET Hamiltonian have already been described in detail elsewhere~\cite{baker2024simulating}. QDET employs the $G_0W_0$ method, which starts from a DFT calculation to obtain the second-quantized effective Hamiltonian
\begin{align}
H_\text{eff} &= \sum_{p,q=1}^N \sum_\sigma t_{pq}^\text{eff}  c_{p\sigma}^\dagger c_{q\sigma} \nonumber \\
& + \frac{1}{2} \sum_{p,q,r,s=1}^N \sum_{\sigma, \sigma^\prime} v_{pqrs}^\text{eff} c_{p \sigma}^\dagger c_{q \sigma^\prime}^\dagger c_{r \sigma^\prime} c_{s \sigma},
\label{eq:H_eff}
\end{align}
where the indices $p,q,r,s$ run over a basis of $N$ single-particle states, $c$ and $c^\dagger$ are respectively the electron annihilation and creation operators, and $\sigma$ denotes the spin quantum numbers. The single-particle states are a selected subset of the Kohn-Sham (KS) orbitals $\phi_n(\bm{r})$ of the defect-containing supercell. The subset of orbitals is typically selected based on its localization factor, defined by~\cite{sheng2022green}
\begin{equation}
L_n = \int_V |\phi_n(\bm{r})|^2 d\bm{r}
\label{eq:loc_factor}
\end{equation}
 This factor is calculated within a volume $V$ enclosing the defect region, and orbitals are selected if their $L_n$ value is below a user-specified threshold. Recently, Otis {\it et al.}~\cite{otis2025strongly} proposed to expand this subset of orbitals by adding KS states below the valence band maximum within an energy window comparable with the band gap of the material. However, for the specific case we consider in \cref{sec:validation}, namely that of the negatively charged boron vacancy in hexagonal boron nitride, the authors found no appreciable change to the defect's electronic structure or excited states with the addition of these extra orbitals. For this reason, in this work we select our subset of KS states using only the localization factor.  
 
 The two-body matrix elements $v_{pqrs}^\text{eff}$ are two-electron integrals of an effective Coulomb interaction $W^E(\bm{r}_1, \bm{r}_2)$ screened by the dielectric response of the host environment~\cite{ma2021quantum, sheng2022green}. The one-body coefficients $t_{pq}^\text{eff}$ are matrix elements of the KS Hamiltonian. These coefficients are corrected by the double-counting term that removes the contributions of the Hartree and exchange-correlation potentials due to the active electrons that had been included in the DFT calculations of the supercell~\cite{sheng2022green}.

Using the same basis of KS states, we define the electric dipole operator~\cite{baker2024simulating} entering~\cref{eq:rad_rate} as
\begin{equation}
\bm{D} = \sum_{p,q=1}^N \sum_\sigma \bm{d}_{pq}^\sigma c_{p\sigma}^\dagger c_{q\sigma},
\label{eq:dipole}
\end{equation}
where $\bm{d}_{pq}$ is the matrix element
\begin{equation}
\bm{d}_{pq}^\sigma = -\int \phi_{p\sigma}^*(\bm{r}) \bm{r} \phi_{q\sigma}(\bm{r}) d\bm{r}
\label{eq:d_me}
\end{equation}
of the electron position operator $\bm{r}$. The matrix elements $\bm{d}^\sigma_{pq}$ are computed numerically on a finite grid of volume $\Delta V$ using the KS wave functions
\begin{equation}
\bm{d}^{\sigma}_{pq} \approx \sum_{\bm{r}_i \in \text{grid}}\phi_{p \sigma}^{\star}(\bm{r}_i)\bm{r}_i\phi_{q \sigma}(\bm{r}_i) \Delta V.
\label{eq:d_pq}
\end{equation}
As we show in~\cref{sec:validation}, the excited states $\ket{E_j}$ with the largest dipole transition amplitudes $|D_{ij}|^2 = |{\bra{E_i}} \bm {D} \ket{E_j}|^2$ determine the dominant features in the absorption or emission spectrum of the quantum defect.

To evaluate the intersystem crossing rate in~\cref{eq:ISC_rate} we use the one-body term of the Breit-Pauli Hamiltonian describing the spin-orbit coupling (SOC) interaction~\cite{marian2012spin}
\begin{equation}
H_\text{SOC} = \frac{\alpha^2}{2} \sum_{i=1}^{N_e} \sum_{I=1}^{N_\text{A}} \frac{Z_I}{|\bm{r}_i-\bm{R}_I|^3} [(\bm{r}_i - \bm{R}_I) \times \bm{p}_i] \cdot \bm{s}_i,
\label{eq:h_soc_op}
\end{equation}
where $N_e$ is the number of electrons populating the KS defect states, and
$N_\text{A}$ is the number of atoms in the supercell. The parameters $Z_I$ and $\bm{R}_I$ are respectively the atomic number and coordinates of the $I$th atom, and $\bm{p}$ and $\bm{s}$ are the one-electron momentum and spin operators. In second quantization the SOC operator in~\cref{eq:h_soc_op} is given by
\begin{equation}
H_\text{SOC} = \sum_{p,q=1}^N \sum_{\sigma\tau} h_{p\sigma,q\tau}^\text{soc} c_{p\sigma}^\dagger c_{q\tau},
\label{eq:soc_op}
\end{equation}
where the matrix elements $h_{p\sigma,q\tau}^\text{soc}$ are defined as
\begin{align}
&\kern -10pt h_{p\sigma,q\tau}^\text{soc} = \frac{\alpha^2}{2} \sum_{\gamma} \int d\bm{r}~\phi_p^*(\bm{r}) \chi_{\sigma}^*(\gamma) \nonumber \\
&\times \left[ \sum_{I=1}^{N_A} \frac{Z_I}{|\bm{r}-\bm{R}_I|^3} [(\bm{r}-\bm{R}_I) \times \bm{p}] \cdot \bm{s} \right]~\phi_q(\bm{r}) \chi_{\tau}(\gamma).
\label{eq:h_soc_me}
\end{align}
In~\cref{eq:h_soc_me} $\chi_{\sigma}(\gamma)$ is the spin function of the $p$th KS orbital $\phi_p(\bm{r})$. The full derivation of the matrix elements is given in~\cref{appx:soc_melements}. Neese showed in Ref.~\cite{neese2005efficient} that the \textit{two-body} terms of the Breit-Pauli Hamiltonian can be included in a mean field approximation to retain the one-body structure of the SOC operator. In this work, we do not consider this contribution to the matrix elements as we have observed they are negligible for our case. However, we note that they can be incorporated straightforwardly into our analysis.

Making the change of variables $\bm{r} \rightarrow \bm{r}+\bm{R}_I$ in~\cref{eq:h_soc_me}, we find
\begin{equation}
\frac{Z_I}{|\bm{r}-\bm{R}_I|^3} [(\bm{r}-\bm{R}_I) \times \bm{p}] \cdot \bm{s} \rightarrow \frac{Z_I}{r^3} \bm{l} \cdot \bm{s},
\label{eq:cvariable}
\end{equation}
where $\bm{l}=\bm{r} \times \bm{p}$ is the angular momentum operator. By using Eqs.~(\ref{eq:ladder1}-\ref{eq:ladder2}) for the ladder operators ($l_{\pm}, s_{\pm}$) we obtain
\begin{equation}
\bm{l} \cdot \bm{s} = \frac{1}{2} (l_+ s_- + l_- s_+) + l_zs_z.
\label{eq:ls_text}
\end{equation}
Note that the first two terms in~\cref{eq:ls_text} flip the spin-projection quantum number $M$, while the third term leaves the spin projection intact (we use the symbol $S$ to refer to total spin). Inserting~\cref{eq:ls_text} into~\cref{eq:h_soc_me} allows us to split the $H_\text{SOC}$ operator as
\begin{equation}
H_\text{SOC}=H_\text{SOC}^z + H_\text{SOC}^\perp,
\label{eq:split_soc}
\end{equation}
where $H_\text{SOC}^z$ drives intersystem crossings between excited states with the same total-spin projections while $H_\text{SOC}^\perp$ couples excited states whose spin quantum numbers differ by $\Delta M = \pm 1$. These observables  can be separately used to compute the axial ($k^{\text{ISC},z}$) and non-axial ($k^{\text{ISC},\perp}$) intersystem crossing rates between the defect excited states, following \cref{eq:ISC_rate}.

For the quantum algorithms below, we will further decompose $H_\SOC$ into its spin tensor operator components. Spin tensor operators $T^{S,M}$ are any operators that fulfill the commutation relations
\begin{align}\label{eq:spin_tensor_relations}
    [s_{\pm}, T^{S,M}] &= \sqrt{S(S+1)-M(M\pm1)} T^{S,M\pm1},\\
    [s_z, T^{S, M}] &= M T^{S,M}.
\end{align}
This gives us a particular way to partition the general SOC operator with respect to its action on the spin sector. The full set of one-body spin tensor operators is of the general form~\cite{helgaker2000}
\begin{equation}
\label{eq:spin_tensor_operators}
\begin{split}
    T^{0,0}_{pq}&= \frac{1}{\sqrt{2}}(c_{p\uparrow}^\dagger c_{q\uparrow} +c_{p\downarrow}^\dagger c_{q\downarrow}),\\
    T^{1,0}_{pq} &= \frac{1}{\sqrt{2}}(c_{p\uparrow}^\dagger c_{q\uparrow} -c_{p\downarrow}^\dagger c_{q\downarrow}),\\
    T^{1,1}_{pq} &= -c_{p\uparrow}^\dagger c_{q\downarrow}, \qquad T^{1,-1}_{pq} =  c_{p\downarrow}^\dagger c_{q\uparrow}.
\end{split}
\end{equation}
The physical meaning of these spin tensor operators is that, owing to their commutation relations, they are responsible for particular transitions between total spin and spin projection sectors. For example, the $T^{0,0}$ spin tensor operator cannot change the total spin nor the spin projection of the state it acts on, since it commutes with both $s_z$ and $s_{\pm}$; by contrast, $T^{1,0}$ can change the total spin by $\Delta S = \pm 1$, but cannot change the spin projection, since it commutes with $s_z$ but not with $s_{\pm}$; and $T^{1,\pm1}$ can change both.  

Since we are interested in isolating particular ISC effects of the SOC operator, it is useful to decompose SOC into spin tensor operators, as this provides more freedom to track separate transition channels individually. Being a one-body operator, $H_\SOC$ can be decomposed into the sum of spin tensor operators $H_\SOC^{S,M}$
\begin{equation}
    H_\SOC = H_\SOC^{0,0} + H_\SOC^{1,0} + H_\SOC^{1,1} + H_\SOC^{1,-1}.
\end{equation}

To determine the form of the elements in the decomposition, we use the definition of the SOC operator and the relations in \cref{eq:spin_tensor_operators}, finding
\begin{align}
    H_\SOC^{1,1} &= \sum_{pq} h_{p\uparrow,q\downarrow}^{\text{soc}} c_{p\uparrow}^\dagger c_{q\downarrow} \nonumber \\
    H_\SOC^{1,-1} &= \sum_{pq} h_{p\downarrow,q\uparrow}^{\text{soc}} c_{p\downarrow}^\dagger c_{q\uparrow} \label{eq:h_soc_decomp}\\
    H_\SOC^{0,0} &= \frac{1}{2} \sum_{pq} \left(h_{p\uparrow,q\uparrow}^{\text{soc}} + h_{p\downarrow,q\downarrow}^{\text{soc}}\right) \left( c_{p\uparrow}^\dagger c_{q\uparrow} + c_{p\downarrow}^\dagger c_{q\downarrow} \right) \nonumber\\
    H_\SOC^{1,0} &= \frac{1}{2} \sum_{pq} \left(h_{p\uparrow,q\uparrow}^{\text{soc}} - h_{p\downarrow,q\downarrow}^{\text{soc}}\right) \left( c_{p\uparrow}^\dagger c_{q\uparrow} - c_{p\downarrow}^\dagger c_{q\downarrow} \right).\nonumber
\end{align}
By summing these operators we can verify that they indeed add up to $H_\SOC$. Further, since the operator part of the expressions has the form given in \cref{eq:spin_tensor_operators}, they respect the commutation relations of \cref{eq:spin_tensor_relations} and thus form a proper spin tensor decomposition of the SOC operator. 

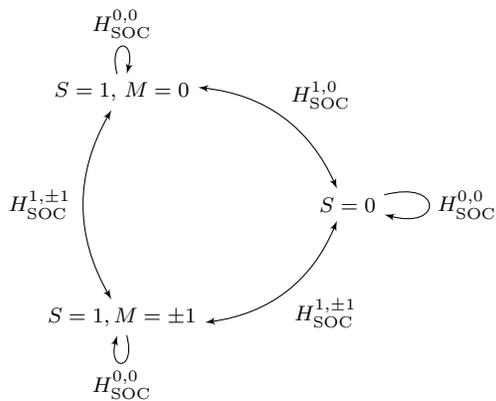
\begin{figure}[t]
     \begin{tikzpicture}[
   every node/.style={font=\footnotesize},
   >=latex, 
   bend angle=30,  
   node distance=2cm
 ]

 \node[] (S1M0) at (0,0) {$S=1$, $M=0$};
 \node[] (S0) at (3,-1.5) {$S=0$};
 \node[] (S1M1) at (0,-3) {$S=1, M=\pm 1$}; 


 S=1, M=0 to S=0
 \draw[latex'-latex', bend left] (S1M0) to node[midway, above right] {$H_\SOC^{1,0}$} (S0);

 \draw[latex'-latex', bend right] (S1M1) to node[midway, below right] {$H_\SOC^{1,\pm 1}$} (S0);

 \draw[latex'-latex', bend left] (S1M1) to node[midway, left=-15pt, align=center, text width=2cm] {$H_\SOC^{1,\pm 1}$} (S1M0);

 \draw[-latex', loop above] (S1M0) to node[midway, above] {$H_\SOC^{0,0}$} ();
 \draw[-latex',  loop right] (S0) to node[midway, right] {$H_\SOC^{0,0}$} ();
 \draw[-latex',  loop below] (S1M1) to node[midway, below] {$H_\SOC^{0,0}$} ();

 \end{tikzpicture}
     \caption{Transitions between spin subspaces enabled by the spin tensor components of the full $H_\SOC$ operator. The definitions of $H_\SOC^{S,M}$ are indicated in~\cref{eq:h_soc_decomp}. We depict $S=1, M=\pm 1$ together despite them being different spin sectors, because no spin tensor from those depicted allows $M$ to change by more than $\pm 1$, so there are no transitions between them.}
     \label{fig:spin_tensor}
 \end{figure}

Having split the SOC operator this way, we have identified terms that are responsible for different spin sector transitions, as depicted visually in~\cref{fig:spin_tensor} for the spin and spin projection state submanifolds $\{S = 0\}$, $\{S = 1, M = 0\}$, and $\{S = 1, M = \pm 1\}$. This will provide more control over which transitions are being simulated by the quantum algorithm. Additionally, we can relate the spin-tensor decomposition to the axial and non-axial components
\begin{align}
    H_\SOC^{z} &= H_{\SOC}^{0,0} + H_{\SOC}^{1,0}, \\
    H_\SOC^{\perp} &= H_{\SOC}^{1,1} + H_{\SOC}^{1,-1}.
\end{align}

\section{Quantum algorithms}
\label{sec:algos}

\begin{table*}[t]
\renewcommand{\arraystretch}{1.6}
\setlength{\tabcolsep}{12pt}
\begin{tabular}{@{} p{2.2cm} | 
                >{\raggedright\arraybackslash}p{\dimexpr(\textwidth - 2.2cm - 6\tabcolsep - 3\arrayrulewidth)/3\relax} | 
                >{\raggedright\arraybackslash}p{\dimexpr(\textwidth - 2.2cm - 6\tabcolsep - 3\arrayrulewidth)/3\relax} | 
                >{\raggedright\arraybackslash}p{\dimexpr(\textwidth - 2.2cm - 6\tabcolsep - 3\arrayrulewidth)/3\relax} @{} }
 & 
 & \multicolumn{2}{c@{}}{\textbf{Inter-system crossing (ISC)}} \\
\cline{3-4}
 & \textbf{Optical response} \newline (\cref{ssec:opt_response})
 & \textbf{Evolution-proxy} \newline (\cref{ssec:comm_based_algo_def})
 & \textbf{Spectroscopy-based} \newline (\cref{ssec:spec_based_algo_def}) \\
\midrule

\textbf{Problem} &
Computing the emission spectrum $\sigma_{S,M}(\omega)$ and assessing optical addressability. &
Detecting ISC rate imbalance ($k^{\text{ISC},z} \neq k^{\text{ISC},\perp}$). &
Detecting ISC rate imbalance \emph{and} identifying the participating eigenstates. \\

\textbf{Description} &
Samples the time-domain Green's function via Hadamard tests and applies a discrete-time Fourier transform to reconstruct the emission spectrum. It also serves as a building block for the spectroscopy-based algorithm. &
Uses short-time evolution overlaps under axial and non-axial SOC components as proxies for the exact ISC rates. Since $H_\SOC$ is a one-body operator, its time evolution can be fast-forwarded in a single Trotter step. &
Compares optical spectra obtained under $H_\eff$ and $H_\eff + \kappa H_\SOC^{1,M}$, where $\kappa$ amplifies the SOC perturbation. Intensity leakage between spin sectors, interpreted via first-order perturbation theory, reveals both the ISC magnitude and the eigenstate origin. \\

\textbf{Advantages} &
Provides the full emission spectrum. &
Lowest gate cost among the three algorithms. &
Identifies which eigenstates drive ISC by locating spectral intensity changes. \\

\textbf{Limitations} &
Does not aim to detect ISC. Achieving high spectral resolution requires long evolution times, and a trade-off between broadening $\eta$ and maximal time steps $j_{\max}$ must be managed. &
Cannot identify which eigenstates participate in ISC, and requires an \textit{a priori} energy cutoff $E'$ for the QSP projective filter. &
Inherits all optical-response errors and adds a first-order perturbation-theory assumption that breaks down if $\kappa$ closes intra-sector energy gaps. Requires three full spectroscopy runs. \\

\bottomrule
\end{tabular}
\caption{Summary of the three quantum algorithms introduced in \cref{sec:algos}, highlighting
their general approach, main advantages, and limitations.
The optical response algorithm (\cref{alg:spectroscopy}) also serves as a subroutine of the spectroscopy-based method.}
\label{tab:algo_comparison}
\end{table*}

As explained at the end of \cref{ssec:odmr_basics}, to detect a nonzero ODMR contrast it is sufficient to focus on detecting whether or not there is an imbalance of axial versus non-axial ISC rates, $k^{\text{ISC},z} \neq k^{\text{ISC},\perp}$. This realization means it is not necessary to precisely calculate the individual rates, as long as there is a robust way of gauging how different they are. This unlocks significant savings from the quantum algorithm perspective, as we now describe. 

\subsection{Evolution-proxy algorithm}
\label{ssec:comm_based_algo_def}

The evolution-proxy algorithm detects the ISC rate imbalance by taking advantage of the fact that the time evolution under $H_\SOC$ at short times is proportional to the exact ISC rate.
Given the model of the low-lying states in \cref{fig:levels_rates}, from \cref{eq:ISC_rate} the axial and non-axial ISC rates are defined as
\begin{align}
    k^{\ISC,z} &= A|\braket{E_{1,S=1, M=0} | H_\SOC^{1,0} |E_{1,S=0}}|^2,\\
    k^{\ISC,\perp} &= A|\braket{E_{1,S=1, M=1} | \left( H_{\SOC}^{1,1} + H_\SOC^{1,-1} \right) |E_{1,S=0}}|^2,
\end{align} 
where $E_{1,S,M}$ is the first excited state in the $S,M$ spin sub-manifold, $A$ collects multiplicative prefactors, and we used the spin tensor decomposition of $H_\SOC$ introduced in \cref{ssec:obs}.
Quantitatively calculating these rates can be costly on a quantum computer. For that reason we instead propose to evaluate proxy quantities $\tilde k^{\perp}(t)$ and $\tilde k^z(t)$, defined via the following time evolution
\begin{align}\label{eq:proxy_rates_nc_a}
    \tilde k^z(t) &= |\braket{E_{1,S=1, M=0} | e^{-i t H_{\SOC}^{1,0}}| E_{1,S=0, M=0}}|^2, \\
    \tilde k^\perp(t) &= |\braket{E_{1,S=1, M=1} | e^{-i t (H_{\SOC}^{1,1} + H_{\SOC}^{1,-1})}|E_{1,S=0, M=0}}|^2.
    \label{eq:proxy_rates_nc_b}
\end{align}
These proxies are proportional to the exact rates over short time intervals, as can be seen by performing a short-time Taylor expansion
\begin{multline}\label{eq:taylor_evolution_proxy}
    \braket{E_{1,S, M} | e^{-it H_\SOC^{S,M} }|E_{1,S=0}} = \\
    - it \braket{E_{1, S=1, M} | H_{\SOC}^{S,M}|E_{1,S=0}} + O(t^3).
\end{multline}
The proxies are thus tightly related to the exact ISC rates. Comparing them should allow us to conclude if $k^{\ISC,z} \neq k^{\ISC,\perp}$. Note that the quadratic error term in \cref{eq:taylor_evolution_proxy} vanishes because the two states in the matrix element belong to different spin sectors, and $H_{\SOC}^{S, M}$ follows the transitions according to~\cref{fig:spin_tensor}. For this reason, the leading order error will be cubic.

We now describe the key stages of the algorithm. \\

\textbf{State preparation:} Evaluating the ISC rate proxies requires preparing the states $\ket{E_{1,S,M}}$---this is also the main challenge in evaluating the ISC rates classically.
Although it is relatively straightforward to prepare a state within a given spin sector, ensuring that the prepared state overlaps only with the desired energy eigenstate or window of interest is challenging. 
Without proper control, high-energy states within the Hilbert space may introduce unwanted contributions to the ISC rates, contaminating the rate comparison.

This challenge is analogous to the ground state preparation problem, where inaccuracies in the classical determination of the ground state can compromise the accurate evaluation of its energy.
For the evolution-proxy algorithm, we first prepare a state with broad support on the spectrum using the dipole operator. Subsequently, we selectively project out undesired high-energy components.
More specifically, we propose the following initial state preparation protocol
\begin{enumerate}
    \item Classically compute the approximate ground state $\ket{\psi_{0,S,M}}$ using advanced classical simulation methods such as density-matrix renormalization group (DMRG).
    \item Apply the dipole operator $\bm{D}$ to it.
    \item Implement the state $\bm{D} \ket{\psi_{0,S,M}}$ in the quantum register, using for example the sum-of-Slaters technique \cite{fomichev2024initial}.
    \item Leverage quantum projective techniques to project in the desired low energy window. To synthesize these projectors we could use a combination of quantum phase estimation with the median lemma to heavily suppress the probability of mis-classification~\cite{nagaj2009fast}. However, we find it more straightforward and cost-effective to leverage (generalized) quantum signal processing~\cite{motlagh2024generalized,berry2024doubling,dong2022ground}, approximating a Heaviside function.
\end{enumerate}
The implementation of QSP is the most cost-intensive step of the state preparation procedure. It will require approximating a Heaviside function with a polynomial of degree inversely proportional to the sharpness of the transition, as a proportion of the total energy window. In this paper, we analyze implementing each step of Hamiltonian evolution via Trotter formulas, described below.

To begin, consider the state $\bm{D}\ket{\psi_{0,S,M}}$ we would like to filter, with support over the energy range $[E_{0,S,M}, E_{\max}]$, where $E_{\max}$ is an upper bound for the largest eigenvalue of $H_\eff$ for which $\bm{D}\ket{\psi_{0,S,M}}$ has support. This upper bound may be obtained, for example, as the energy of the most high-energy single Slater determinant entering into $\bm{D}\ket{\psi_{0,S,M}}$. We then choose the energy window $[E_{0,S,M}, E']$ aligned with the low-energy subspace we are interested in filtering around. The situation is illustrated in \cref{fig:QSP_filter_conceptual}. The filtering window cutoff $E'$ should be chosen to encompass the eigenstate that is expected to participate in the ISC transitions. For example, in the diagram of \cref{fig:levels_rates}, these would be the first excited states of the singlet and triplet sectors. An approximate determination of this cutoff may be accomplished by first applying the optical response algorithm (see the following \cref{ssec:opt_response}) to find the energies of strongly radiatively active states. Alternatively, a guess may be available \textit{a priori} from experimental measurements. Together, the eigenvalue window over which the initial state has support and the filtering cutoff set the evolution time of a single evolution step in quantum signal processing. As a consequence, calling $\lambda:= \max(E_{\max}-E', E'-E_{0})$, the evolution time and degree of the quantum signal processing filter are rescaled by $1/\lambda$, as we explain next.

There are two key parameters in quantum signal processing: the width of the transition area $\Delta$ and the precision with which we implement the polynomial, $\epsilon_{\text{QSP}}$. The former must be normalized in such a way that the eigenvalue range spans $[-1, +1]$. This can be done renormalizing  $\Delta \rightarrow \Delta/\lambda$, see~\cref{fig:QSP_filter_conceptual}. The degree of the polynomial required will be $d = O\left(\frac{\lambda}{\Delta}\text{poly}\log (\epsilon_{\text{QSP}}^{-1})\right)$, where $\epsilon_{\text{QSP}}$ represents the error in the polynomial fitting of the Heaviside function outside of the $\Delta/\lambda$ region. An analytical numerical upper bound can be found in Theorem 17 in~\cite{pocrnic2025constant}.

\begin{figure}[t]
    \centering
    \includegraphics[width=1.0\linewidth]{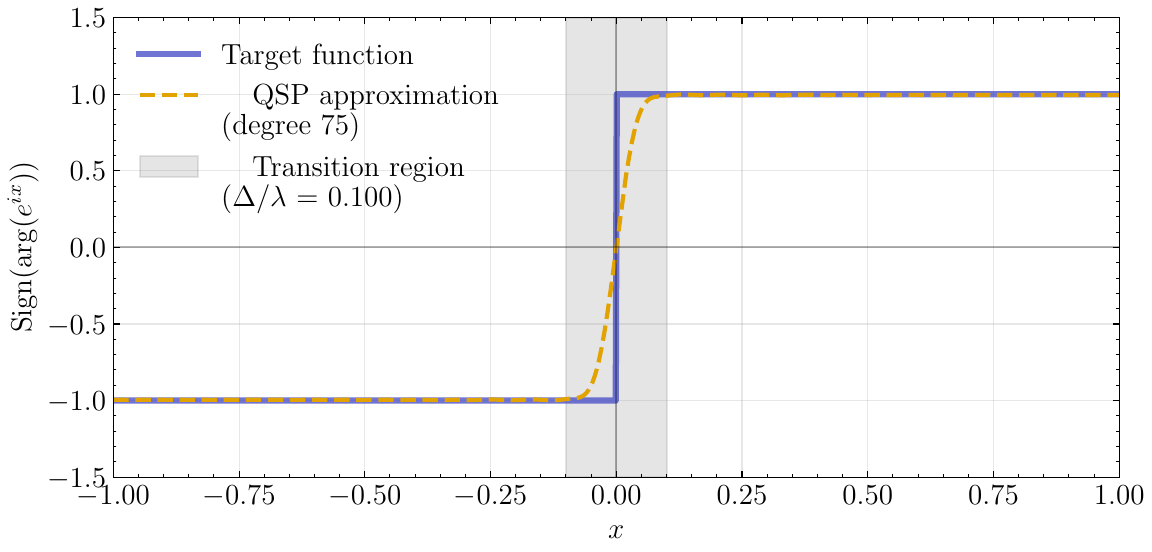}
    \caption{Quantum signal processing polynomial approximating a Heaviside function, syntesized with \texttt{pyqsp}~\cite{mrtc_unification_21,cdghs_finding_qsp_angles_20, haah_decomposition_19, gslw_qsvt_19, dmwl_efficient_phases_21}. The Heaviside function can be used to probabilistically project an ancilla to $\ket{0}$ or $\ket{1}$ depending on the sign of the energy of the eigenstate.}
    \label{fig:Heaviside}
\end{figure}

\textbf{Time evolution:} While in principle any method may be used to implement the time evolution $\exp(-i H_\eff)$, we choose an approach based on qubit-efficient product formulas coupled with an optimized representation of the Hamiltonian through compressed double factorization (CDF) \cite{fomichev2025affordable,cohn2021quantum}. In this section we give a concise description following Ref.~\cite{fomichev2025affordable}. Given $H_\eff$ as defined in \cref{eq:H_eff}, we can apply the CDF ansatz to factorize the one- and two-electron integrals as 
\begin{align}
    t_{pq}^\eff =& \sum_k \tilde{U}_{pk}^{(0)} \tilde{Z}_{kk}^{(0)} \tilde{U}_{qk}^{(0)},\label{eq:cdf-onebody} \\
   v^\eff_{pqrs} \approx & \sum_{\ell=1}^L \sum_{k,l=1}^N U^{(\ell)}_{pk} U^{(\ell)}_{qk} Z^{(\ell)}_{kl} U^{(\ell)}_{rl} U^{(\ell)}_{sl} \label{eq:cdf-twobody}.
\end{align}
\begin{figure}
\definecolor{XanaduBlue}{HTML}{4D53C8}
\definecolor{XanaduRed}{HTML}{D7333B}
\definecolor{XanaduGreen}{HTML}{52BD7C}
\definecolor{XanaduOrange}{HTML}{E2A300}
    \centering
    \begin{tikzpicture}[scale=0.75]

    \draw[->] (-5,0) -- (1,0) node[above] {$\mathbb{R}$}; 
    \draw[thick,XanaduBlue] (-4,0) -- (-3,0);
    \draw[thick,XanaduOrange] (-3,0) -- (0,0);
    \foreach \x/\l in {-4/$E_0$,-3/$E'$,0/$E_{\max}$}
        \draw (\x,0.1) -- (\x,-0.1) node[below] {\l};

    \draw (3,0) circle (1cm);
    \draw[thick,XanaduBlue] (2,0) arc (180:240:1cm); 
    \draw[thick,XanaduOrange] (2,0) arc (180:0:1cm);
    \node at (4,1) {$\ket{0}$};
    \node at (4,-1) {$\ket{1}$};

    \draw[dashed] (1.75,0) -- (4.25,0); 

\end{tikzpicture}
    \caption{Assume $\bm{D}\ket{E_0}$ has support over $[E_0, E_{\max}]$, and we are interested in projecting into an energy window $[E_0, E']$. We can pad the smaller segment to make them of equal size. Then, we can project both energy ranges to 0 and 1, so that a single qubit suffices to identify whether the state belongs to the desired energy window or is outside of it. }
    \label{fig:QSP_filter_conceptual}
\end{figure}
Here the matrices $U^{(\ell)}$ are special orthogonal matrices representing single-particle basis rotations, and $Z^{(\ell)}$ are symmetric matrices. In the case of the one-body integrals, these matrices may be obtained through a direct diagonalization of the matrix $t_{pq}^\eff$, with the eigenvectors giving $\tilde U^{(0)}$ and eigenvalues allowing to construct the diagonal matrix $\tilde Z^{(0)}$. In the case of the two-electron integrals, this is accomplished approximately, by solving an optimization problem that finds such $U^{(\ell)}$ and $Z^{(\ell)}$ that minimize the difference between the left and right hand sides of \cref{eq:cdf-twobody}. Once these matrices are found, they may be used to re-write the Hamiltonian by transforming the single-particle creation and annihilation operators
\begin{equation}
    a^{(\ell)\dagger}_{k\gamma} = \sum_{p} U_{pk}^{(\ell)} a^\dagger_{p\gamma}, \quad a^{(\ell)}_{k\gamma} = \sum_{q} U_{qk}^{(\ell)} a_{q\gamma},
\end{equation}
and applying the Jordan-Wigner transformation in the simple form
\begin{equation}
    \hat{n}_{k\tau} = \frac{1 - \sigma_{z,k\tau}}{2}.
\end{equation}
The Hamiltonian may then be written directly in terms of the $Z^{(\ell)}$ and the $U^{(\ell)}$ matrices. The  $Z^{(\ell)}$ matrices give the coefficients of single and pair-wise Pauli $Z$ rotations. Meanwhile $\bm{U}^{(\ell)}$ are the full Hilbert space unitaries that modify a given many-body state in response to changing the single-particle basis by a $U^{(\ell)}$ rotation. The full Hamiltonian takes the form
\begin{align}
    H &= \left(E + \sum_k Z_k^{(0)} -\frac{1}{2}\sum_{\ell,kl}Z_{kl}^{(\ell)}+\frac{1}{4} \sum_{\ell,k} Z^{(\ell)}_{kk} \right)\bm{1}\nonumber\\ \label{eq:CDF_Hamiltonian}
    &-\frac{1}{2}\bm{U}^{(0)} \left[\sum_k Z^{(0)}_{k} \sum_\gamma  \sigma_{z,k\gamma} \right](\bm{U}^{(0)})^T\\
    &+\frac{1}{8}\sum_\ell  \bm{U}^{(\ell)} \left[\sum_{(k,\gamma)\neq(l,\tau)}  \left( Z^{(\ell)}_{kl} \sigma_{z,k\gamma} \sigma_{z,l\tau}\right) \right](\bm{U}^{(\ell)})^T.
\end{align}
Time evolution under this Hamiltonian may be implemented using standard Trotter product formulas. The general idea is to continuously perform basis rotations $\bm{U}^{(\ell)}$ to pass to a basis where a given fragment of the Hamiltonian is diagonal, implement the corresponding qubit rotations, then pass to the next basis. In this work, we primarily employ the second-order Trotter product formula. The unitaries $\bm{U}^{(\ell)}$ may be constructed using Thouless's theorem \cite{kivlichan2018quantum,thouless1960stability} and implemented using Givens rotations \cite{arrazola2022universal}, and the Pauli $Z$ rotations can be readily compiled to the Clifford + $T$ gateset. 

The number of Trotter steps required may be determined using a perturbation-theory approach as presented in Ref.~\cite{loaiza2025simulating,martinez2023estimating}, and applied to the electronic structure Hamiltonian in the context of spectroscopy in Ref.~\cite{fomichev2025affordable}. The key idea is to observe that product formulas implement \textit{exact} time evolution under an \textit{approximate} Hamiltonian. The time step and the order of the product formula determine the degree of approximation of the Hamiltonian. For a Hamiltonian written as a sum of fragments $H = \sum_i H_i$, a second-order product formula may be shown via the Baker-Campbell-Hausdorff formula to implement
\begin{align}
    U_2(h) = \exp(-ih [H_\eff -h^2 \hat Y_3 + h^4 \hat Y_5 +\ldots])
\end{align}
where $Y_{2k+1}$ is the error operator, given by linear combinations of nested commutators of the Hamiltonian's individual fragments. For example, for a second order product formula the leading order is
\begin{align}
    \hat Y_3&= \sum_j\left( - \frac{\left[\left[\sum_{i<j}H_{i},H_j\right],\sum_{i<j}H_{i}\right]}{12}\right)\nonumber \\
    &+ \sum_j\left(- \frac{\left[\left[\sum_{i<j}H_{i}, H_j\right],H_j\right]}{24}\right).
\end{align}
From this perspective, to control the Trotter error we can demand that the difference in eigenvalues between the true Hamiltonian $H_\eff$ and the approximate Hamiltonian $\tilde H_\eff \approx H_\eff + h^2 \hat Y_3$ implemented by the second-order product formula is less than a specific accuracy cutoff $\epsilon_{\text{HS}}$. This cutoff may be taken to be the expected broadening $\eta$ in the defect's optical response. The eigenvalue difference may be estimated perturbatively as 
\begin{equation}
     E_{n,S,M} - \tilde E_{n,S,M} \approx h^2 \bra{E_{n,S,M}} \hat Y_3 \ket{E_{n,S,M}}.
\end{equation}
We call
\begin{equation}
    \epsilon_{\text{HS}} :=\max_{n,S,M} \, \bra{E_{n,S,M}} \hat Y_3 \ket{E_{n,S,M}}.
\end{equation}
We take $h^2\epsilon_{\text{HS}}$ similar to the broadening $\eta$ in the spectroscopy algorithm; or $\Delta$ in the evolution-proxy algorithm.

\textbf{Time evolution by $H_\SOC$:} A key advantage of turning the ISC rate problem into one of time-evolving a given state is that we avoid the rather expensive qubitization or block-encoding of $H_{\SOC}$. Such encoding is necessary for a direct rate calculation. Moreover, since $H_{\SOC}$ is a one-body operator, time evolution under it can be fast-forwarded. Specifically, we apply a basis rotation to diagonalize it 
\begin{equation}
  h_{p\sigma,q\tau}^{\text{soc}} = U_0 Z_0 U_0^\dagger, \quad Z_0 = \mathrm{diag}(\lambda_1, \dots, \lambda_{2N}).
\end{equation}
where $U_0$ is the single-particle basis transformation matrix and $\{\lambda_{p\sigma}\}$ are the eigenvalues of $H_{\SOC}^{S, M}$. Thus implementing the time evolution step is straightforward via
\begin{equation}
    e^{it H_\SOC^{S,M}} = \bm{U}_0 \prod_{p\sigma} e^{it \lambda_{p\sigma} \sigma_{z,p\sigma}} \bm{U}_0^\dagger
\end{equation}
where $\bm{U}_0$ is the unitary that rotates the many-body state within the full Hilbert space space induced by the single-particle basis rotation $U_0$, that can be constructed and implemented as above, and $\sigma_{z,p\sigma}$ are Pauli $Z$ operators for the $(p,\sigma)$ orbital. \\

\textbf{Obtaining the matrix element:} Having described how to perform time evolution and to prepare the required states, we now explain the way in which we evaluate the matrix element. There are two measurement schemes for doing so: the swap test, and a modified Hadamard test. 
Each scheme has its own advantages and disadvantages, making it suitable for different scenarios. 

The modified Hadamard test, as depicted in \cref{fig:modified_Hadamard}, is a particular way to measure the \textit{off-diagonal} matrix element of the unitary controlled by the Hadamard ancilla (see \cref{appx:modified_Hadamard} for details)~\cite{parrish2019quantum}. Unlike the standard Hadamard test, this circuit can measure arbitrary matrix elements, not just expectation values. This is accomplished through a non-trivial state preparation step. The sum-of-Slaters technique and subsequent QSP projection prepare an initial state $\ket{\zeta_{t = 0}}$:
\begin{equation}
\ket{\zeta_{t=0}}_s = \frac{\alpha}{\sqrt{2}\gamma}\ket{0}\ket{\psi_{S=1,M}}_s + \frac{\beta}{\sqrt{2}\gamma}\ket{1}\ket{\psi_{S=0}}_s.    
\end{equation}
This state is then used in the familiar Hadamard test with the unitary $e^{-it H_{\SOC}^{1, M}}$, and now yields the off-diagonal matrix element of the unitary with respect to $\ket{\psi_{S=1,M}}_s$ and $\ket{\psi_{S=0}}_s$. The modified Hadamard test can be advantageous when dealing with small ISC rate proxies, as it avoids squaring the already small value. It also eliminates the need for an additional register. However, the result depends on the amplitudes of the prepared states within the target energy window, which need to be estimated separately---for example, through the swap test, or by using amplitude estimation.

\begin{figure}[t]
\[
    \Qcircuit @C=1em @R=.7em {
  \ket{0} &&\qw & \multigate{1}{\text{SoS}} & \gate{S^\dagger} & \ctrl{1} &  \gate{H} &\meter \gategroup{1}{5}{1}{5}{.7em}{--}  \\
  \ket{0}_s &&{/}\qw & \ghost{\text{SoS}} & \multigate{1}{\text{QSP}}  & \gate{e^{-it H_{\SOC}^{1, M}}}  &\qw &\qw&  \\
  \ket{0} && \qw{/} & \qw & \ghost{\text{QSP}} & \meter & &   &  & \\ 
    }
\]
    \caption{Implementation of the modified Hadamard test~\cite{parrish2019quantum}. The sum-of-Slaters technique~\cite{fomichev2024initial} and subsequent Quantum Signal Processing projection and postselection prepare the state 
    $
        \ket{\zeta_{t=0}}_s = \frac{\alpha}{\sqrt{2}\gamma}\ket{0}\ket{\psi_{S=1,M}}_s + \frac{\beta}{\sqrt{2}\gamma}\ket{1}\ket{\psi_{S=0}}_s,
    $
    see~\cref{appx:modified_Hadamard} for a detailed discussion.
     When the $S^\dagger$ gate is present (respectively absent), the circuit measures the imaginary (real) component of the matrix element~\cref{eq:proxy_rates_nc_a,eq:proxy_rates_nc_b}.
    }
    \label{fig:modified_Hadamard}
\end{figure}

An alternative to the modified Hadamard test is the usual swap test, see~\cref{fig:swap_test} and \cref{appx:swap}. The swap test's advantage is its independence from state preparation success probabilities. However, it is more resource-intensive than the modified Hadamard test for achieving high precision with small overlaps, and also necessitates an extra qubit register.

Having described all the key aspects of the approach, we summarize the evolution-proxy algorithm in a step-by-step description presented in \cref{alg:evolution_proxy}.

\begin{algorithm}[H]
\caption{Evolution-proxy algorithm}
\label{alg:evolution_proxy}
\begin{algorithmic}[1]
    \State Classically compute $\ket{\psi_{0,S=0}}$ as well as $\ket{\psi_{0,S=1,M=0}}, \ket{\psi_{0,S=1,M=1}}$, where $\ket{\psi_{0,S,M}}$ is the approximate ground state in the corresponding $\{S,M\}$ manifold.
    \State For each spin sector, apply the dipole operator $\bm{D}$ and classically remove the ground-state contribution by subtracting the expectation value of the dipole operator
    \Statex \parbox[t]{\dimexpr\linewidth-2em}{%
        \begin{equation}\label{eq:psi_forallE}
            \ket{\psi_{\mathbf{D}, S, M}} = \bm{D}\ket{\psi_{0,S,M}} - \braket{\psi_{0,S,M}|\bm{D}|\psi_{0,S,M}}\ket{\psi_{0,S, M}}.
        \end{equation}}
    \State Use the sum-of-Slaters technique to prepare approximations $\ket{\psi_{\mathbf{D}, S, M}}$~\cite{fomichev2024initial} in the appropriate registers.
    \State Use Quantum Signal Processing (QSP) to probabilistically project these states into the low energy subspace $[E_0, E']$. This effectively implements the projector
    \Statex \parbox[t]{\dimexpr\linewidth-2em}{%
        \begin{align}
            \ket{\psi_{1,S, M}}:=\left(\sum_{E<E'}\ket{E}\bra{E}\right)\ket{\psi_{\mathbf{D}, S, M}}.
        \end{align}}
    \State Select the appropriate spin tensor components $H_{\SOC}^{S, M}$ of the full SOC operator as written in \cref{eq:proxy_rates_nc_a,eq:proxy_rates_nc_b}.
    \State Fast forward $\ket{\psi_{1,S=0}}$ under the chosen spin tensor components of SOC via a single Trotter step.
    \State Use the modified Hadamard test (or the swap test) to compute the rate proxies
    \Statex \parbox[t]{\dimexpr\linewidth-2em}{%
        \begin{align}\label{eq:tilde_k_z}
            \tilde k^z(t) &= \braket{\psi_{1,S=1, M=0} | e^{-it H_{\SOC}^{1, 0}}|\psi_{1,S=0}},\\ \label{eq:tilde_k_perp}
            \tilde{k}^\perp(t) &= \braket{\psi_{1,S=1, M=1} | e^{-it \left( H_{\SOC}^{1,1} + H_{\SOC}^{1,-1} \right)}|\psi_{1,S=0}}.
        \end{align}}
    \State Evaluate the rate proxy ratio $\tilde k^{\ISC,z} / \tilde k^{\ISC,\perp}$ and deduce whether $k^{\ISC,z} \neq k^{\ISC,\perp}$.
\end{algorithmic}
\end{algorithm}

\subsection{Optical response of the defect}
\label{ssec:opt_response}
Our second proposed algorithm for determining the ISC rate is based upon computing the defect's emission spectrum with and without the SOC operator. For this reason, in this section we first briefly summarize the quantum algorithm to obtain the emission spectrum of the spin defect, originally proposed in Ref.~\cite{fomichev2024simulating} for X-ray spectroscopy and adapted here for spin defects. Even beyond detecting ODMR activity, such an algorithm is useful in the search for better spin defects for quantum sensing. Spin defects for sensing applications must have optically-addressable excited states and large emission rates. Experimentally, observing sharp and intense peaks in the optical absorption/emission spectrum of the defect is a key indicator of optical addressability and large emission rates. Being able to simulate the spectrum thus allows to directly assess the suitability of candidate defects for sensing applications.

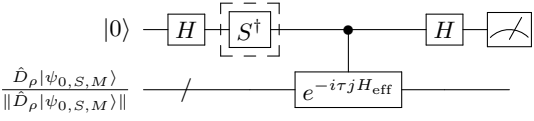
\begin{figure}[t]
    \centering
    \[
    \Qcircuit @C=1em @R=1em {
    \lstick{\ket{0}} & \gate{H} & \gate{S^\dagger} & \ctrl{1} & \gate{H} & \meter \gategroup{1}{3}{1}{3}{.7em}{--} \\
    \lstick{\frac{\hat{D}_\rho\ket{\psi_{0,S,M}}}{\|\hat{D}_\rho\ket{\psi_{0,S,M}}\|}} & {/}\qw & \qw & \gate{e^{-i\tau j H_\eff}} & \qw & \qw \\
    }
    \]
    \caption{Hadamard test circuit to compute the time-domain Green's function in \cref{eq:G(t)}. After preparing the normalized state $\hat{D}_\rho\ket{\psi_{0,S,M}}/\|\hat{D}_\rho\ket{\psi_{0,S,M}}\|$, we perform the Hadamard test on the unitary for time evolution under the system Hamiltonian. Adding (removing) the phase gate $S^{\dagger}$ gives the real (imaginary) part of $\tilde{G}(\tau j)$.}
    \label{fig:algo-circs-timedomain}
\end{figure}

The target quantity of the algorithm is the emission cross-section $\sigma(\omega)$. By Fermi's golden rule, within a given spin sector it is given by
\begin{equation}
    \sigma_{S,M}(\omega) = \sum_{n \neq 0} \sum_{\rho = x, y, z}\frac{\left| \bra{E_{n,S,M}} D_\rho \ket{E_{0,S,M}} \right|^2 \eta}{((E_{n,S,M} - E_{0,S,M}) - \omega)^2 + \eta^2},
    \label{eq:crosssection}
\end{equation}
where $E_{n,S,M}$ is the $n$th excited state in the $\{S,M\}$ spin manifold with corresponding energy $E_{n,S,M}$, $D_\rho$ is the $\rho$th Cartesian component of the dipole operator $\bm{D}$, and $\eta$ is the broadening (typically due to a combination of finite experimental resolution and finite state lifetimes). 
The overlap $\bra{E_{n,S,M}} D_\rho \ket{E_{0,S,M}}$ is precisely the dipole transition amplitude between the $n$th excited state and the ground state entering the definition of the radiative decay $k^r$ in~\cref{eq:rad_rate}.
It can be shown \cite{fomichev2024simulating} that up to an additive constant, this cross-section may be written as the following discrete-time Fourier transform Green's function
\begin{equation}
    \sigma_{S,M}(\omega) \approx \frac{\tau}{2\pi} \sum_{\rho=x,y,z} \sum_{j=-j_{\text{max}}}^{j_{\text{max}}} e^{-\eta\tau|j|} \tilde{\mathcal{G}}_\rho(\tau j) e^{ij\tau\omega},
    \label{eq:green-via-discretetime-fourier}
\end{equation}
with time step $\tau \sim O(1/\|\hat H\|_\omega)$, where $\|H\|_\omega$ is the support of the dipole operator in the Hamiltonian spectrum, a maximal evolution time $\tau j_{\text{max}}$, and with the time-domain Green's function $\tilde{\mathcal{G}}_\rho(\tau j)$ at time $\tau j$ given by
\begin{equation}\label{eq:G(t)}
    \tilde{\mathcal{G}}_\rho(\tau j) = \bra{E_{0,S,M}} D_\rho e^{-i \tau j  H_\eff } D_{\rho} \ket{E_{0,S,M}}.
\end{equation}
These matrix elements can be estimated using the Hadamard test (\cref{fig:algo-circs-timedomain}), with the time evolution operator $\exp(-i \tau j H_\eff)$ implemented using CDF and Trotter product formulas, exactly as described in detail for the previous algorithm. Parameters such as the time step $\tau$, total shot budget $S_\text{Had}$ and its allocation among the $j$ times, and the maximal evolution time $\tau j_{\text{max}}$, may be fixed as described in Ref.~\cite{fomichev2025affordable}. The preparation of the approximate initial state $\ket{\psi_{0,S,M}} \approx \ket{E_{0,S,M}}$ may be done similarly to the description in the preceding \cref{ssec:comm_based_algo_def}, namely by a combination of pre-computing approximately the ground state classically, and preparing that state using the sum-of-Slaters method. 

We summarize the algorithm for obtaining the optical response of the defect in a step-by-step form in \cref{alg:spectroscopy}. 

\begin{algorithm}[H]
\caption{Defect optical response algorithm
\label{alg:spectroscopy}}
\begin{algorithmic}[1]
        \State Classically pre-compute $\ket{\psi_{0,S,M}}$, the approximate ground state of the system. 
        \State Classically apply the dipole operator $\bm{D}$ while removing the ``self-absorption'' contribution by subtracting the expectation value of the dipole operator
        \begin{equation}
            \ket{\psi_{\bm{D},S, M}} = \bm{D}\ket{\psi_{0,S, M}} - \braket{\psi_{0,S, M}|\bm{D}|\psi_{0,S, M}}\ket{\psi_{0,S, M}}.
        \end{equation}
        \State Use the sum-of-Slaters method to prepare $\ket{\psi_{\bm{D},S, M}}$~\cite{fomichev2024initial} in the system register.
        \State Use time evolution and Hadamard tests to compute $\tilde{\mathcal{G}}(\tau j) = \braket{\psi_{\bm{D},S, M}|e^{-i \tau j \hat H_\eff}|\psi_{\bm{D},S, M}}$, for $j\in(-j_{\max}, j_{\max})$. See Ref.~\cite{fomichev2025affordable} for details on how to set the cutoff $j_{\max}$ and allocate the shot budget $S_\text{Had}$ among the different times $\tau j$, as well as for other optimizations in the implementation.
        \State Use the expression \cref{eq:green-via-discretetime-fourier} to obtain the spectrum.
\end{algorithmic}
\end{algorithm}

\subsection{Spectroscopy-based algorithm}\label{ssec:spec_based_algo_def}

Having described the algorithm for obtaining the emission spectrum of a given defect (\cref{alg:spectroscopy}), here we show how to use it to detect an ISC rate imbalance and thus ODMR activity. The main idea behind this \textit{spectroscopy-based} algorithm is to obtain and compare the optical response of the system by performing time evolution under $H_\eff$ and under $H_\eff + \kappa H_\SOC$, for a multiplicate prefactor $\kappa$. If there is non-negligible ISC occurring due to $H_\SOC$, spectral intensity will transfer to adjacent spin sectors. This transfer manifests as reductions in dipole-driven intensity compared to the spectrum obtained using only $H_\eff$. From these changes in the spectrum, we may deduce the magnitude of the ISC rate between different spin sectors, as well as identify the participating states. More rigorously, we will use perturbation theory to demonstrate how $H_\SOC$ leads to dipole intensity leakage from the spectrum. Application of perturbation theory to this problem is enabled by the fact that the energy scale of $H_{\SOC}$ is typically much smaller than that of the electronic Hamiltonian $H_\eff$. 

Using (non-degenerate) perturbation theory, the new eigenstates $\ket{E'_{n,S,M}}$ and eigenvalues $E'_{n,S,M}$ of the perturbed Hamiltonian $H' = H_\eff + \kappa H_\SOC$ are given by \cite{Zwiebach18Perturbation}
\begin{multline}
    \ket{E'_{n,S,M}} = \ket{E_{n,S,M}} \\
    - \kappa \sum_{\substack{m\neq n,\\ S',M'}} \frac{\braket{E_{m,S',M'}|H_{\SOC}|E_{n,S,M}}}{E_{m,S',M'}-E_{n,S,M}}
    \ket{E_{m,S',M'}}\\
    +O((\kappa H_{SOC})^2).
    \label{eq:perturbed_eigenstates_HSOC}
\end{multline}
\begin{multline}
    E'_{n,S,M} = E_{n,S,M}  +\kappa\braket{E_{n,S,M}|H_{\SOC}|E_{n,S,M}}\\
    -\kappa^2 \sum_{\substack{m\neq n,\\S',M'}} \frac{|\braket{E_{m,S',M'}|H_{\SOC}|E_{n,S,M}}|^2}{E_{m,S',M'}-E_{n,S,M}}+ O((\kappa H_{SOC})^3).
    \label{eq:perturbed_eivalues_HSOC}
\end{multline}
Here $\kappa = 1$ is the exact case of just adding the $H_\SOC$ operator directly. However, a scaling factor $\kappa$ can be used to artificially increase the magnitude of this perturbation. This makes its spectral effects easier to detect. Such amplification must be carried out within limits, so as not to entirely scramble the electronic structure of the original Hamiltonian. If there is degeneracy in the original Hamiltonian $H_\eff$ that is lifted by $H_{\SOC}$, we obtain similar results by using degenerate perturbation theory, see~\cref{appx:degenerate}. This modification of the eigenstates and eigenvalues by the presence of $H_\SOC$ has a direct and observable impact on the spectrum, which within a given spin manifold $\{S,M\}$ is now given by
\begin{align}\label{eq:perturbed_spectrum}
    \sigma_{S,M}(\omega) = \sum_{n \neq 0} \sum_{\rho = x, y, z}\frac{\left| \bra{E'_{n,S,M}} D_\rho \ket{E_{0,S,M}} \right|^2 \eta}{((E'_{n,S,M} - E_{0,S,M}) - \omega)^2 + \eta^2}.
\end{align}

In~\cref{eq:perturbed_eigenstates_HSOC}, we observe that the sum over eigenstates runs over all spin sectors. Later down this section will substitute $H_{SOC}$ by the components of $H_{SOC}$ that changes quantum numbers $S, M$, using the decomposition in~\cref{fig:spin_tensor}.
Thus, if there is a state $\ket{E_{m,S',M'}}$ relatively close in energy to $\ket{E_{n,S,M}}$ but with different spins $S \neq S', M \neq M'$, such that $\braket{E_{m,S',M'}|H_\SOC|E_{n,S,M}} \neq 0$, then \cref{eq:perturbed_eigenstates_HSOC} effectively shows that the modified eigenstate $\ket{E'_{n,S,M}}$ will have a non-negligible contribution from a state \textit{outside the original spin manifold} $\{S,M\}$. Since $\braket{E_{0,S,M}|D_\rho|E_{m,S',M'}} = 0$ because the dipole operator does not change spin character, by normalization of $\ket{E'_{n,S,M}}$ we must have
\begin{equation}
|\braket{E_{0,S,M}|D_\rho|E'_{n,S,M}}|^2<|\braket{E_{0,S,M}|D_\rho|E_{n,S,M}}|^2.
\end{equation}
Through \cref{eq:perturbed_spectrum}, this means the corresponding intensity of the $E_{n,S,M}$ peak is reduced, with the reduction proportional to the ISC rate,  as seen in \cref{eq:perturbed_eigenstates_HSOC}. Moreover, since we can observe the modified peak's energy position, we can deduce which eigenstate is participating in ISC---additional information that we could not gather from the evolution-proxy algorithm.  

This demonstrates how to detect the ISC rate through its effect on the spectrum in principle. However, in practice due to the significant disparity in energy scales between $H_\eff$ and $H_\SOC$, this might be difficult to observe, or require long evolution times and thus be costly. This is where the $\kappa$ prefactor proves useful. It can be increased to artificially boost the strength of the spin-orbit coupling perturbation. Since the fundamental goal is to detect an ISC rate \textit{imbalance} rather than precisely estimate each rate, as long as both $k^{\ISC,z}$ and $k^{\ISC,\perp}$ are boosted by the same amount, this will not affect the relative rate comparison. We then choose a large enough $\kappa$ so the perturbation of $H_\SOC$ to the spectrum is visible, but not so large that it dramatically modifies the eigenstates of $H_\eff$. Rigorously, the effective restriction on the size of $\kappa$ is that it cannot be so large that it closes the smallest energy gap between eigenstates within the same spin sector. More precisely, using the perturbation theory results for eigenvalues in \cref{eq:perturbed_eivalues_HSOC}, the choice of $\kappa$ must satisfy
\begin{equation}\label{eq:kappa_size}
    \kappa |\braket{E_{n,S,M}|H_\SOC|E_{n,S,M}}|< \min_{m} |E_{n,S,M} - E_{m,S,M}|.
\end{equation}
While the right-hand side of this can be difficult to estimate, a straightforward upper bound to the left-hand side may be obtained by diagonalizing $H_\SOC$, giving a method of choosing $\kappa$.

We now describe step-by-step how to implement the spectroscopy-based algorithm for the particular problem of detecting ISC rate imbalance in a spin defect. As in the general spectroscopy algorithm, the first step is to prepare an approximate initial state. For both the axial and non-axial rates, we will begin by preparing an approximate singlet ground state $\ket{\psi_{0,S=0}}$. We then apply the optical response algorithm as written (\cref{alg:spectroscopy}), with the only modification being that we select specific spin tensor components of $H_\SOC$ to be used during time evolution, depending on which rate channel is of interest
\begin{itemize}
    \item for the axial rate, $H_\eff + \kappa H_\SOC^{1,0}$,
    \item for the non-axial rate, $H_\eff + \kappa (H_\SOC^{1,1} + H_\SOC^{1,-1})$. 
\end{itemize}    
In addition to being able to separate the channels, a key advantage of using spin tensor components is reducing the energy shift of the eigenstates due to the artificially enhanced perturbation strength. This can be seen from \cref{eq:perturbed_eivalues_HSOC}, where for any spin tensor component of the form $H_{\SOC}^{1, M}$
the leading order perturbation to the energy, $\braket{E_{n,S,M}|H_{\SOC}^{1, M}|E_{n,S,M}}$ vanishes, since $H_{\SOC}^{1, M}$ does not commute simultaneously with $S^2$ and $S_z$. This allows using stronger perturbations, getting ISC-driven changes to be more visible while making only minor modifications to the electronic structure. Having obtained the spectra after finishing the application of \cref{alg:spectroscopy}, we then visually compare them with the reference spectra obtained without $H_\SOC$, observe the spectral differences and identify the energies of the states most active in this process. We then compare those differences between the axial and non-axial case to determine one of them is more pronounced, thus deducing the presence or absence of an ISC rate imbalance and hence ODMR activity.

Lastly, the spectroscopy-based algorithm, summarized in \cref{alg:spectroscopy-iscrates}, can be modified substituting the spin-tensor operator components $H_{\text{SOC}}^{1,M}$ with the commutator $[H_{\text{SOC}}, S^2]$. Evolving under the commutator accomplishes the same task of separating out the non-spin-conserving part of $H_\SOC$. Although this substitution is not expected to offer an advantage in our problem, it could prove beneficial when investigating different symmetries or symmetries where the tensor decomposition is not feasible. The exponential of commutators can be simulated with the methods in~\cite{childs2013commutators,chen2022commutators,casas2025approximating}, while the implementation of $\exp(-i S^2)$ is explained in~\cref{app:S^2}.

\begin{algorithm}[H]
\caption{Spectroscopy-based algorithm
\label{alg:spectroscopy-iscrates}}
\begin{algorithmic}[1]
        \State Classically pre-compute $\ket{\psi_{0,S=0}}$, the approximate singlet ground state of the system for the Hamiltonian $H_\eff$. 
        \State Classically apply the dipole operator $\bm{D}$ while removing the ``self-absorption'' contribution by subtracting the expectation value of the dipole operator
        \begin{equation}
            \ket{\psi_{\bm{D},S=0}} = \bm{D}\ket{\psi_{0,S=0}} - \braket{\psi_{0,S=0}|\bm{D}|\psi_{0,S=0}}\ket{\psi_{0,S=0}}.
        \end{equation}
        \State Use the sum-of-Slaters method to cheaply prepare $\ket{\psi_{S, M}}$~\cite{fomichev2024initial} in the system register.
        \State Obtain a reference spectrum by using \cref{alg:spectroscopy} on $H_\eff$.
        \State For $\kappa$, obtain modified spectra by using \cref{alg:spectroscopy} on
        \begin{itemize}
            \item $H_\eff + \kappa H_{\SOC}^{1,0}$ for the axial rate, 
            \item $\hat H_\eff + \kappa (H_{\SOC}^{1,1} + H_{\SOC}^{1,-1})$ for the non-axial rate.
        \end{itemize}
        \State Compare the spectra against the reference and identify spectral leakage. 
        \State Compare the spectral leakage between the axial and non-axial channels and qualitatively deduce the amount of ISC rate imbalance present.
\end{algorithmic}
\end{algorithm}

\section{Resource estimation}
\label{sec:resource_estimation}
Having introduced two quantum algorithms for detecting an ISC rate imbalance and one for probing the optical response of a spin defect, in this section we analyze their computational costs, specifically the number of logical qubits and non-Clifford gates required, while accounting for factors such as Trotter error, sampling complexity, and state preparation. By combining detailed analysis with empirical data, we aim to provide a comprehensive understanding of the quantum resources necessary for practical implementation.

\subsection{Cost of the evolution-proxy algorithm}

The overall number of Toffoli gates needed to execute the evolution-proxy algorithm can be directly computed from the algorithm's key steps, namely 
\begin{enumerate}
    \item Preparing the initial states $\ket{\psi_{\bm{D},S,M}}$ with the sum-of-Slaters algorithm, a cost we denote by $C_\text{SOS}(D)$ for a combination of $D$ Slater determinants;
    \item Using Quantum Signal Processing (QSP) to filter out high-energy components from those states. The filter is approximated with a $d = O(\lambda/\Delta \times \text{poly}\log (\epsilon_{\text{QSP}}^{-1}))$ degree polynomial~\cite{motlagh2024generalized,berry2024doubling}. Each call to the oracle implemented through Trotter-based time evolution with per-step cost $C_{\text{Trot}}(H_\text{eff})$ for $N_{\text{Trot}}(h^2\epsilon_{\text{HS}})\times d$ steps, where $N_{\text{Trot}}(h^2\epsilon_{\text{HS}})$ is the number of Trotter steps to control the Trotter error to a target $\Delta = h^2\epsilon_{\text{HS}}$.
    \item Time-evolving by a spin tensor component of the $H_\SOC$ operator with cost denoted by $C_{H_\SOC^{S,M}}$;
    \item Performing the modified Hadamard test measurement $S_\text{Had}(\epsilon)$ times to achieve precision $\epsilon$ of the proxy quantity $\tilde k^{z}, \tilde k^{\perp}$. 
\end{enumerate}
The cost of all these steps may be combined in the overall expression
\begin{align}\label{eq:cost_evolution_proxy}
   & C_{\text{ev-pr}} = 2\times 2\times S_{\text{Had}}(\epsilon) \times \\
   &\left[ \frac{C_{\text{SOS}}(2D)+ C_{\text{Trot}}(H_\text{eff}) \times N_{\text{Trot}}(\Delta)\times d}{\gamma^2} + C_{H_{\SOC}^{S, M}}\right].\nonumber
\end{align}
Here the factors of $2$ out front are coming from the need to measure the real and imaginary parts of the proxy rate, as well as obtain both proxy quantities $\tilde k^{z}$ and $\tilde k^{\perp}$. The first term $C_{\text{SOS}}(2D)$ inside the square brackets represents the cost of preparing the state 
\begin{equation}
    \ket{\zeta_{t=0}} = \frac{\alpha}{\gamma\sqrt{2}}\ket{0}\ket{\psi_{\bm{D}, S =1, M}}+\frac{\beta}{\gamma\sqrt{2}}\ket{1}\ket{\psi_{\bm{D}, S =0}}
\end{equation}
required by the modified Hadamard test as in \cref{fig:modified_Hadamard} using the sum-of-Slaters method (this cost is given in Ref.~\cite{fomichev2024initial}); the second term is the cost of QSP; and the third term $C_{H_{\SOC}^{S, M}}$ is the cost of the controlled fast-forwarding of an SOC spin tensor component. The factor $\gamma^2$ is the average of $\alpha^2$ and $\beta^2$.

To employ the cost equation above, we need to answer three key questions: (i) how many Hadamard test repetitions $S_\text{Had}(\epsilon)$ are required to achieve the target precision, (ii) what target failure probability $p_\text{fail}$ should be used for state preparation, and (iii) what is the cost of implementing the Hamiltonian simulation $C_\text{Trot}(H_{\text{eff}})\times N_{\text{Trot}}(\Delta)\times d$ within QSP and the cost $C_{H_{\SOC}^{S,M}}$ of fast-forwarding $H_{\SOC}^{1, M}$. 

The Hadamard test sampling complexity is 
obtained by using error bounds from the binomial distribution to estimate the number of measurements for a target sampling error $\epsilon$. The probability $p(0)$ of measuring on the ancilla qubit is used to estimate the real (imaginary) expectation value $\mu$ as
\begin{equation}
    \mu = p(0)-p(1) = 2p(0) - 1.
\end{equation}
Thus, the error $\sigma(p(0))$ propagates to the quantity of interest $\mu$ as $2\sigma(\mu)$ for each of the real and imaginary component, totalling $\epsilon = 4\sigma(\mu)$. Overall, this implies that
\begin{align}
    \sigma^2\left(\mu\right) = \frac{p(1-p)}{S_{\text{Had}}}\leq \frac{1}{4S_{\text{Had}}},
\end{align}
where the worst case obtains for $p=1/2$.
Thus, the total number of measurements should be at least
\begin{equation}\label{eq:Hadamard_shots}
    S_{\text{Had}}(\epsilon) \geq \frac{1}{4\sigma^2(\mu)} = \frac{4}{\epsilon^2}\left(\frac{\gamma^2}{\alpha\beta}\right)^{2}.
\end{equation}
The factor $\gamma^2 / \alpha\beta$ is needed to rescale $\epsilon^{-1}$ and recover the desired matrix element to the original precision $\epsilon$ in the \textit{modified} Hadamard test (see~\cref{appx:modified_Hadamard} for a more detailed derivation). This overhead may be reduced using amplitude estimation~\cite{brassard2002quantum}. In that case we would only require~\cite{grinko2021iterative} 
\begin{multline}
    S_{\text{Had}}(\epsilon) \approx \frac{1.6}{\sigma(\mu)}\log\left(\frac{2}{c}\log_2\left(\frac{\pi}{4\sigma(\mu)}\right)\right)\\
    \approx \frac{6.4}{\epsilon}\frac{\gamma^2}{\alpha\beta}\log\left(\frac{2}{c}\log_2\left(\frac{\pi}{\epsilon}\frac{\gamma^2}{\alpha\beta}\right)\right),
    \label{eq:iterative_AE}
\end{multline}
where $1-c$ is the probability of the measurement being closer than $\epsilon$ to the true $p(0)$.

When the state preparation fails to correctly flag the correct energy window in which we projected, which happens with probability $\epsilon_{\text{QSP}}^2$, this affects the result of the expectation value $\mu$ in the same way as the case described above of having a finite number of samples. Thus, $\epsilon_{\text{QSP}}$ acts as the precision $\epsilon$, limiting the accuracy of the Hadamard test. However, in contrast to $\epsilon$ the number of oracle calls needed to decrease $\epsilon_{\text{QSP}}$ grows only polylogarithmically with $\epsilon_{\text{QSP}}^{-1}$. The reason is that, as described above, the degree of the polynomial and the number of calls to the quantum walk operator scale as $d = O(\frac{\lambda}{\Delta}\text{poly}\log (\epsilon_{\text{QSP}}^{-1})))$. 

Finally, the cost of time evolution in QSP depends also on the cost of a Hamiltonian simulation oracle call $C_{\text{Trot}}\times N_{\text{Trot}}(\Delta)$. To implement one step of the second-order Trotter product formula that we primarily use in this work, we require two first-order Trotter steps. For each of those steps, we need to implement, for each fragment, a single unitary $\bm{U}^{(\ell)}$, which can be implemented with $2N^2$ Pauli rotations, and also a single $Z^{(\ell)}$ matrix, requiring $N(2N+1)$ Pauli rotations ($2N$ for the one-body term). Each Pauli rotation can be synthesized with $\log_2(\epsilon_{\text{rot}}^{-1})$ Toffoli gates \cite{gidney2018halving}. Similarly, fast-forwarding $H_{\SOC}^{1, M}$ incurs a cost equivalent to that of implementing time evolution by the one-body operator. The control on these rotations, coming from the Hadamard test, may be removed using the `double phase trick'~\cite{fomichev2025affordable}, where each $R_z(\theta)$ rotation is sandwiched between two CNOT gates. This approach minimizes the circuit depth while maintaining accuracy, directly impacting the overall resource estimates. Beyond this, we also exploit the other optimizations for Trotter-based time evolution presented in Ref.~\cite{fomichev2025affordable}, including combining consecutive rotations and tuning rotation precision to the application. 

While $N_{\text{Trot}}(\Delta)$ is a function of the Trotter simulation accuracy, the polynomial degree $d$ dictates the QSP filter failure probability and transition width. Both of these parameters directly influence the sharpness of the boundary between the two possible projection outcomes in~\cref{fig:QSP_filter_conceptual}. The Trotter error is estimated as described in \cref{ssec:comm_based_algo_def}, namely by employing an empirical approach based on perturbation theory, evaluating $\epsilon_{\text{HS}} \approx \max \bra{E_{n,S,M}} \hat Y \ket{E_{n,S,M}}$ for a number of trial eigenstates $\ket{E_{n,S,M}}$ for smaller systems that are within reach of simulation, and extrapolating to larger systems. With these estimates in hand, the Trotter step $h$ is chosen so as to keep the overall $\epsilon_{\text{HS}}$ below a user-specified threshold. The number of Trotter steps in that case will grow proportional to $h^{-1} \leq \sqrt{\epsilon_{\text{HS}} / \Delta}$.

\subsection{Optical response and spectroscopy-based ISC quantum algorithm}

In this section we analyze the cost of the optical response and spectroscopy-based algorithm for ISC rates. The cost of performing \cref{alg:spectroscopy} can be estimated as \cite{fomichev2025affordable}
\begin{equation}\label{eq:cost_spectroscopy_initial}
    C_{\text{spec}} = 3\cdot 2\cdot C_{\text{Trot}} \sum_{j=1}^{2j_{\max}}  \, N_{\text{Trot},j} \cdot S_{\text{Had},j},
\end{equation}
where $C_{\text{Trot}}$ is the cost of a single Trotter step, $N_{\text{Trot},j}$ the number of Trotter steps needed to evolve the system for time $\tau j$, and $S_{\text{Had},j}$ the number of shots needed to estimate the real and imaginary component (factor of 2) of matrix element $\tilde{\mathcal{G}}(\tau j) = \braket{\psi_{\bm{D},S,M}|e^{-i\tau j H_{\text{eff}}}|\psi_{\bm{D},S,M}}$, across each of the three Cartesian axes (factor of 3). The cost of a CDF Trotter step can be calculated as described in previous sections, with the sole difference being that $H_\SOC$'s basis rotations affect both spins concurrently. 

There are four relevant sources of error that affect the cost, two related to the implementation of the quantum circuit and two more due to the discrete-time Fourier transform. (The current section summarizes the results derived in Ref.~\cite{fomichev2025affordable}.) One of them is the Trotter error, which we already described how to control in \cref{ssec:comm_based_algo_def}, and which allows to determine the Trotter step size $h$ and thus the number of Trotter steps $S_{\text{Had},j} = \tau j / h$ for any time $j$. 

The second source of error to highlight is the finite sampling in each Hadamard test. As shown in~\cite{fomichev2025affordable} we should take a total of
\begin{align}
    S_{\text{Had}} = \left(\frac{\eta \tau \sum_{k=1}^{2j_{\max}} e^{-|k|\tau \eta}}{2\pi\epsilon_{\text{meas}}}\right)^2
\end{align}
shots to achieve an error $\epsilon_{\text{meas}}$. The shots should be distributed according to the probability distribution 
\begin{equation}
S_{\text{Had},j} = S_{\text{Had}}\frac{e^{-|j|\tau \eta}}{\sum_{k=1}^{2j_{\max}}e^{-\tau\eta |k|}}.
\end{equation}
The third source of error comes from truncating the range of $j$s at which $\tilde{\mathcal{G}}(j\tau)$ is sampled from $j\in (-\infty, +\infty)$ to $j\in(-j_{\max}, +j_{\max})$. This error can be bounded as
\begin{align}
    \epsilon_{\text{Trunc}} \leq \frac{e^{-|j_{\max}\tau\eta|}}{\pi\tau\eta}.
\end{align} 
Finally, the fourth error that arises due to the discrete nature of the discrete-time Fourier transform may be estimated as~\cite{fomichev2025affordable}
\begin{align}
    \epsilon_\tau &\leq \frac{(2j_{\max}\tau)^3}{48(2j_{\max})^2}\max_{j}\left|\frac{d^2}{dt^2}\braket{\psi_{S,M}|e^{-i t (H-w-i\eta)}|\psi_{S,M}}\right|_{t =j\tau}\nonumber\\
    &\leq \frac{j_{\max}\tau^3}{24}\left|H-w-i\eta\right|^2.
\end{align}
Thus, we can take $j_{\max} = \tilde{O}(\tau^{-1}) = \tilde{O}(\|H\|_w)$, for $\|H\|_w$ an energy window that is the size of the support of the dipole operator in the Hamiltonian spectrum~\cite{fomichev2025affordable}. 

With these choices for $\tau$ and $\eta$,~\cref{eq:cost_spectroscopy_initial} can be rewritten as
\begin{align}\label{eq:N2T}
        C_{\text{spec}} 
        &= 3 \cdot 2\cdot C_{\text{Trot}}\cdot s \cdot \frac{\tau}{h} \sum_{j=1}^{2j_{\text{max}}} \frac{ j  \exp{(-\tau\eta |j|)}}{\sum_{k=1}^{2j_{\max}} e^{-|k|\tau \eta}} ,
\end{align}
from which the cost of computing a spectrum for a given Hamiltonian may be directly evaluated.

\section{Application: boron vacancy in hexagonal boron nitride}
\label{sec:validation}

In this section we apply the developed quantum algorithms in a proof-of-concept demonstration to evaluate a well-studied spin defect---a negatively charged boron vacancy in hexagonal boron nitride (hBN)---for suitability for quantum sensing. After discussing the construction of the effective Hamiltonian for this system using the QDET embedding theory, we demonstrate that the algorithms perform as expected by running them on a PennyLane \texttt{lightning} simulator \cite{bergholm2018pennylane,asadi2024} for small system sizes and comparing the results to the classically computed exact reference. Having thus benchmarked the algorithms, we perform a resource estimation for different sizes of the effective Hamiltonian, demonstrating that obtaining the optical response and gauging ODMR activity of this defect is possible with only a few hundred qubits and around a billion Toffoli gates.

\subsection{Defining the system}
\begin{figure*}[t]
\definecolor{XanaduBlue}{HTML}{4D53C8}
\definecolor{XanaduRed}{HTML}{D7333B}
\definecolor{XanaduGreen}{HTML}{52BD7C}
\definecolor{XanaduOrange}{HTML}{E2A300}
    \centering
    \begin{minipage}[b]{0.31\textwidth}
        \begin{tikzpicture}
            \node[anchor=south west, inner sep=0] (image) at (0,0) {\includegraphics[width=\linewidth]{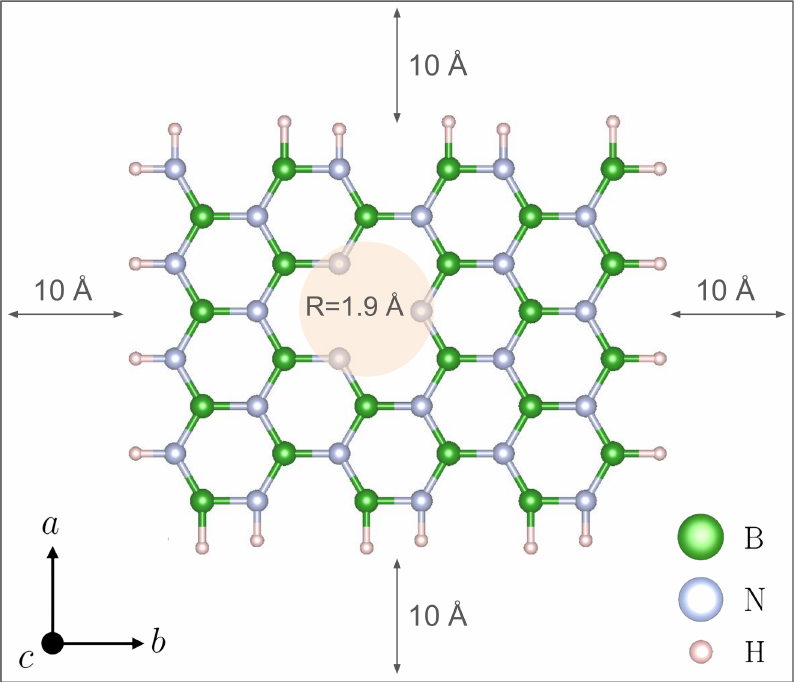}};
            \node[anchor=north west] at (image.north west) {\textbf{a)}};
        \end{tikzpicture}
    \end{minipage}
    \hfill
    \begin{minipage}[b]{0.33\textwidth}
        \begin{tikzpicture}
            \node[anchor=south west, inner sep=0] (image) at (0,0) {\includegraphics[width=\linewidth]{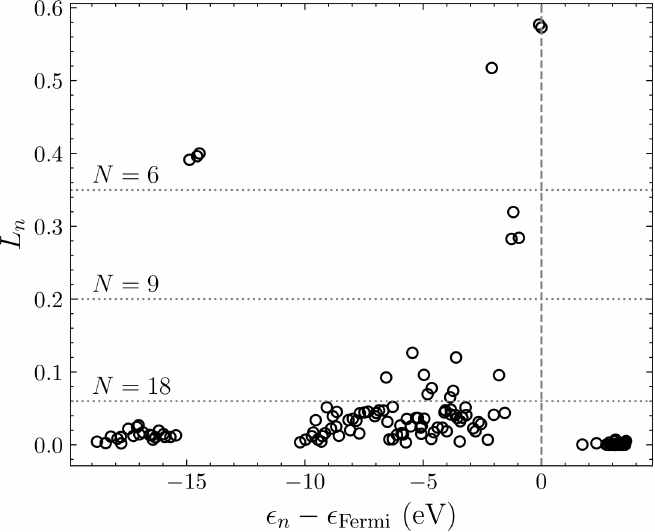}};
            \node[anchor=north west] at (0.75, 4.75) {\textbf{b)}};
        \end{tikzpicture}
    \end{minipage}
    \hfill
    \begin{minipage}[b]{0.33\textwidth}
        \begin{tikzpicture}[scale=0.65]
             \node[anchor=north west] at (0.1, 6.5) { \textbf{c)}};

             \draw[thick, -] (0,0) -- (7,0);
             \draw[thick, -] (0,0) -- (0,6.5);
             \draw[thick, -] (7,6.5) -- (0,6.5);
             \draw[thick, -] (7,6.5) -- (7, 0);

             \foreach \x in { 2,  5.5}
               \draw (\x,0) -- (\x,-0.2);

             \foreach \y in {0.0, 0.5, 1.0, 1.5, 2.0, 2.5, 3.0, 3.5, 4.0}
               \draw (0,1.5*\y+0.2) -- (-0.2,1.5*\y+0.2) node[left, font=\scriptsize] {\y};

             \node[font=\scriptsize] at (2,-0.6) {$S=1$};
             \node[font=\scriptsize] at (5.5,-0.6) {$S=0$};

             \node[rotate=90, anchor=center, font=\scriptsize] at (-1.5, 3.25) {Energy (eV)};

             \draw[thick] (1, 0.2) -- (3, 0.2);
             \draw[thick] (1, 0.2+1.5*3.597) -- (3, 0.2+1.5*3.597);
             \draw[thick] (1, 0.2+1.5*3.797) -- (3, 0.2+1.5*3.797);

             \draw[thick] (4.5, 0.2+1.5*0.920) -- (6.5, 0.2+1.5*0.920);
             \draw[thick] (4.5, 0.2+1.5*1.048) -- (6.5, 0.2+1.5*1.048);
             \draw[thick] (4.5, 0.2+1.5*3.351) -- (6.5, 0.2+1.5*3.351);
             \draw[thick] (4.5, 0.2+1.5*3.788) -- (6.5, 0.2+1.5*3.788);

             \path (1.75, 0.2) coordinate (A);
             \path (1.75, 0.2+1.5*3.797) coordinate (B);
             \draw[thick, dashed, <->, color = XanaduBlue] (A) -- (B) node[midway, left, font=\tiny] {3.80 eV};

             \path (2.25, 0.2) coordinate (A);
             \path (2.25, 0.2+1.5*3.597) coordinate (B);
             \draw[thick, dashed, <->, color = XanaduBlue] (A) -- (B) node[midway, right, font=\tiny] {3.60 eV};

             \path (5.5, 0.2+1.5*0.920) coordinate (A);
             \path (5.5, 0.2+1.5*3.351) coordinate (B);
             \draw[thick, dashed, <->, color = XanaduRed] (A) -- (B) node[midway, left, font=\scriptsize] {2.43 eV};

             \path (3, 0.2+1.5*3.597) coordinate (A);
             \path (4.5, 0.2+1.5*3.351) coordinate (B);
             \draw[thick, dashdotted, <->, color = XanaduOrange] (A) -- (B) node[midway, above, font=\scriptsize] {$k^{\text{ISC}}_z$};
             \draw[thick, dashdotted, <->, color = XanaduOrange] (A) -- (B) node[midway, below, font=\scriptsize] {$k^{\text{ISC}}_\perp$};
        \end{tikzpicture}
    \end{minipage}

    \caption{a) Structural model of a negatively charged boron vacancy $\text{V}_\text{B}^-$ in a hexagonal boron nitride (hBN) cluster $\text{B}_{23}\text{N}_{24}\text{H}_{20}$. A sphere with radius 1.9 \AA{} centered at the vacancy is used to quantify localization of Kohn-Sham states in the defect region. b) Localization factor $L_n$ of the Kohn-Sham orbitals computed within the sphere. The energy $\varepsilon_n$ of the orbitals are given relative to the Fermi energy $\varepsilon_{\text{Fermi}}$. The labels $N=6, 9, 18$ are, respectively, the number of KS orbitals with localization factor $L_n > 0.35, 0.20, 0.06$. c) Low-lying spectra of the many-body defect electronic states obtained by exact diagonalization of the QDET Hamiltonian built using $N=9$ Kohn-Sham orbitals. Dashed vertical lines indicate the radiative transitions, the dot-dashed line shows the axial and non-axial ISC rates $k^{\text{ISC},z}$ and $k^{\text{ISC},\perp}$ between spin sublevels $|E_{1,S=1,M=0}\rangle$ and $|E_{1,S=1,M=\pm 1}\rangle$ of the first triplet excited state and the third singlet state $|E_{2,S=0}\rangle$.}
    \label{fig:model-loc-spec}
\end{figure*}

To construct the boron vacancy in hBN, we begin by optimizing the geometry of a 67-atom two-dimensional hexagonal boron nitride cluster with the vacancy at its center (see \cref{fig:model-loc-spec}a). We perform DFT calculations using an orthogonal supercell with primitive vectors $a=L(1, 0, 0), b=L(0, 1, 0)$, $c=L(0, 0, 0.5)$, where the lattice constant $L=32$ $\text{\AA}$ includes a vacuum layer of $10$ $\text{\AA}$ to avoid artificial interaction between periodic images. Simulations were run with the \texttt{QUANTUM ESPRESSO} package~\cite{Giannozzi2020} using the PBESol~\cite{Perdew2008} density functional and the optimized norm-conserving Vanderbilt pseudopotentials~\cite{Hamann2013}. The geometry of the cluster was relaxed until ionic forces were less than $10^{-3}$ eV/\AA{}.
  
The Kohn-Sham (KS) states at the $\Gamma$ point were calculated using a restricted open-shell occupation in the spin triplet configuration. In \cref{fig:model-loc-spec}b we show the orbital localization factor $L_n$ calculated in a sphere with radius 1.9 $\text{\AA}$ centered at the boron vacancy. The energies are reported relative to the Fermi energy $\varepsilon_\text{Fermi}$ marked at zero. For a given threshold value of $L_n$ we select a set of orbitals to construct the effective QDET Hamiltonian, the dipole $\bm{D}$, and spin-orbit coupling $H_\text{SOC}$ operators for quantum simulations. For example, for $L_n>0.35$ we have $N=6$ localized orbitals. As shown in~\cref{fig:model-loc-spec}b, more states can be selected by gradually decreasing the threshold value.

 \begin{table}[b]
\caption{Excitation energies and dipole transition amplitudes for the defect excited states for the $N=9$ QDET effective Hamiltonian. $E_0$ is the ground-state energy within each total-spin sector.}
\label{table:diptimes}
\centering 
\begin{tabular}{c c c}
\toprule 
& $~~~E_i-E_0$ (eV) & $~~~~~|{\bra{E_i}} \bm {D} \ket{E_0}|^2$ (a.u.) \\ 
\midrule 
& $3.597$ & $5.887$ \\
$S=1$ & $3.781$ & $0.004$ \\
& $3.797$ & $5.726$ \\
\midrule 
 & $0.128$ & $1.997$  \\
$S=0$ & $2.431$ & $4.928$   \\
& $2.869$ & $0.0$   \\
\bottomrule 
\end{tabular}
\end{table}

\begin{table}[b]
\caption{Matrix elements of the SOC operators between optically active low-lying states, in $\text{cm}^{-1}$.}
\label{table:me_soc}
\centering 
\begin{tabular}{c c c}
\toprule 
 & $\bra{E_{1,S=1, M=0}} \hat{O} \ket{E_{2,S=0}}$ & $\bra{E_{1,S=1, M=1}} \hat{O} \ket{E_{2,S=0}}$ \\ 
\midrule 
$H^{\perp}_\SOC$ & 0 & $0.066 + 0.005i$ \\
\\
$H^{z}_\SOC$     & $7 \times 10^{-12}$ & $0$ \\
\bottomrule 
\end{tabular}
\end{table}

The QDET Hamiltonian within this subspace of localized orbitals is built using the \texttt{WEST} code~\cite{govoni2015large, yu2022gpu}. For benchmarking the algorithms, we perform full configuration interaction (FCI) calculations to diagonalize the Hamiltonian in different sectors of the total-spin projection quantum number $M=0, 1$. This is done by using different reference states. For example, for 
$N=9$ localized states, we use the DFT occupation in which the two orbitals closest to the Fermi energy are singly occupied, while the remaining lower-energy orbitals are doubly occupied. This defines a reference state with $M=1$ as we have 9 spin-up and 7 spin-down electrons (9$\uparrow$, 7$\downarrow$). Alternatively, a closed-shell reference with 8 spin-up and 8 spin-down electrons (8$\uparrow$, 8$\downarrow$) was used to compute the defect states with $M=0$. In addition, we calculated the expectation value of total-spin operator $S^2$ to identify the eigenstates with spin $S=1$ and $S=0$.

The low-lying eigenstates of the defect Hamiltonian represented using $N=9$ active orbitals are shown in~\cref{fig:model-loc-spec}c. We used the FCI eigenstates to compute the dipole transition amplitudes $\vert D_{i0}\vert^2=\vert \bra{E_i}\bm{D} \ket{E_0}\vert^2$ to identify the optically-active excited states. The transitions determining the absorption/emission peaks in the triplet and singlet channels have been indicated in~\cref{fig:model-loc-spec}c with vertical dashed lines. For completeness, the calculated energies and dipole transition amplitudes are also reported in \cref{table:diptimes}.

To detect ODMR activity, the goal is to compare the axial and non-axial ISC between the lowest-lying \textit{optically active} states of the defect. If neither an external magnetic field nor zero-field splitting are considered, each state with $S=1$ is a triplet manifold of spin sublevels with quantum numbers $M=0, \pm 1$. Thus, the axial and non-axial ISC rates are determined respectively by matrix elements
\begin{align}
k^{\ISC, z} &= |\braket{E_{2,S=0}|H^{z}_\SOC|E_{1,S=1, M=0}}|^2, \\
k^{\ISC,\perp} &= |\braket{E_{2,S=0}|H^{\perp}_\SOC|E_{1,S=1, M=1}}|^2,
\end{align}
between the first optically active states in the singlet and triplet sectors.~\cref{table:me_soc} shows the ISC rates between the many-body triplet and singlet states computed classically. By definition, the axial operator $H^{z}_\SOC$ does not allow ISC between states with different total-spin projections $M$ while the non-axial operator $H^{\perp}_\SOC$ couples transitions between states whose spin-projections differ by $\Delta M = \pm 1$. Importantly, we observe that the allowed axial and non-axial matrix elements are quite different, which is a necessary condition to observe an ODMR contrast.

\subsection{Simulating the algorithms}

In the previous section we constructed the effective Hamiltonian and the dipole and SOC operators for the boron vacancy in hBN. We also calculated the dipole intensities, thus identifying the optically active low-lying excited states, and estimated the ISC rates for this spin defect. Using the Hamiltonian and the dipole and SOC operators, we now simulate our proposed quantum algorithms for the system $N = 9$ to check their performance against the classically obtained reference values of excited-state energies, dipole intensities, and ISC rates. All algorithms are implemented using PennyLane \cite{bergholm2018pennylane} exactly as described in \cref{sec:algos} and simulated with the \texttt{lightning} backend \cite{asadi2024}.

\begin{figure}
    \centering
    \includegraphics[width=0.7\columnwidth]{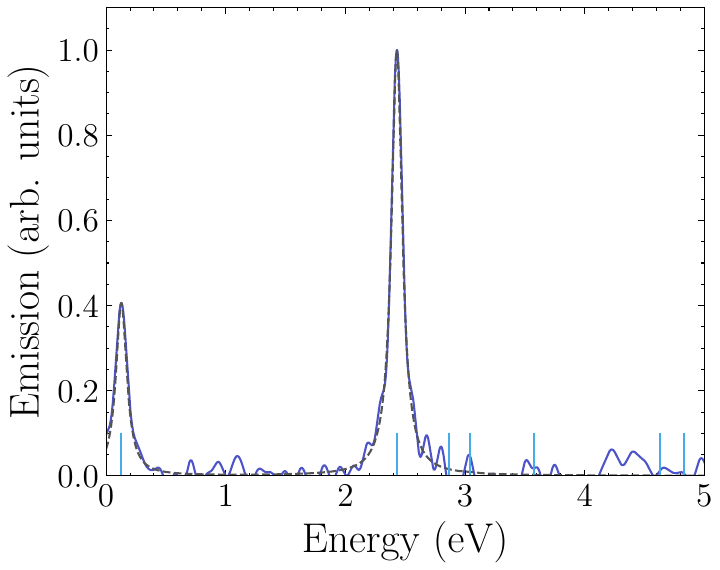}
    \includegraphics[width=0.7\columnwidth]{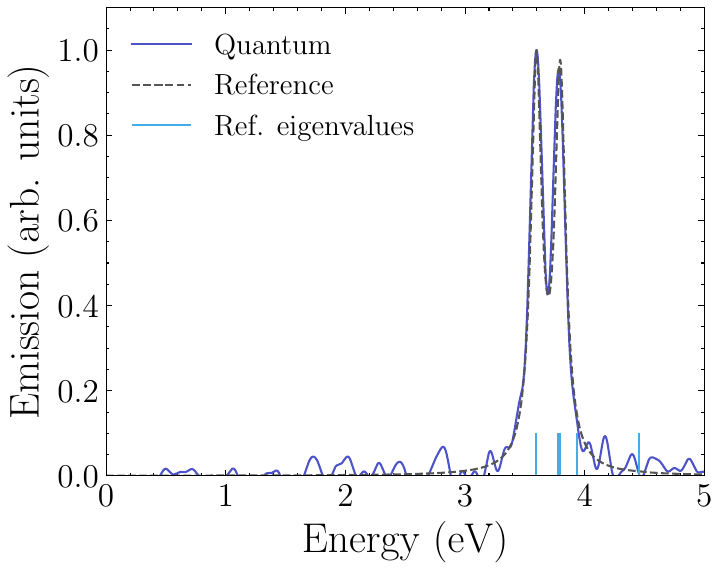}
    \caption{Simulated emission spectra of $\text{V}_\text{B}^-$ in hBN in the (top) singlet ($S=0$) and (bottom) triplet ($S=1$) sectors using an effective Hamiltonian built with $N=9$ orbitals. We use a second order product formula and take $s = 3000$ shots, $\eta=2\cdot 10^{-3}$ Ha, $\|H\|_w=1$ Ha, $h= \tau = \pi/(2\|H\|_w)$, $j_{\max}=500$. The energies of the defect excited states are superimposed using vertical (blue) lines. We observe that the spectroscopy algorithm reproduces the optical response to good accuracy comparing to the results reported in~\cref{table:diptimes}.}
    \label{fig:optical_response}
\end{figure}

First we simulate the optical response algorithm for the boron vacancy in an active space of $N = 9$ spatial orbitals---which amounts to a 19 qubit simulation---deploying a second order product formula with $S_{\text{Had}} = 3000$ shots, broadening of $\eta=2\cdot 10^{-3}$ Ha, effective dipole support window $\|H\|_w=1$ Ha, Trotter step of size $h = \tau = \pi / (2\|H\|_w)$, and maximal evolution time $j_{\max}=500$. In the CDF algorithm we use the number of fragments $L = \tilde{O}(N)$ to make the error constant across system sizes. This empirical rule of thumb is commonly applied in the literature~\cite{motta2021low,von2021quantum,lee2021even}, and we also test it in our specific system (see~\cref{fig:cdf_error} in~\cref{app:empirics}). Specifically, we select $L = N$, which provides mHa accuracy in the simulations. We also select a precision $\epsilon_{\text{rot}} = 10^{-4}$ in the single-qubit rotations, corresponding to a similar precision in the individual values of the CDF approximation. As shown in~\cref{fig:rotation_error} in~\cref{app:empirics}, this leads to a similar target accuracy, which remains close to the mHa range. The results of simulating the algorithm are shown in~\cref{fig:optical_response}, compared against a classical reference. We find that the excitation energies and dipole intensities are captured in exact agreement with the classical reference. 

\begin{figure}[t]
    \centering
    \begin{tikzpicture}
        \node (img1) {\includegraphics[width=0.7\columnwidth]{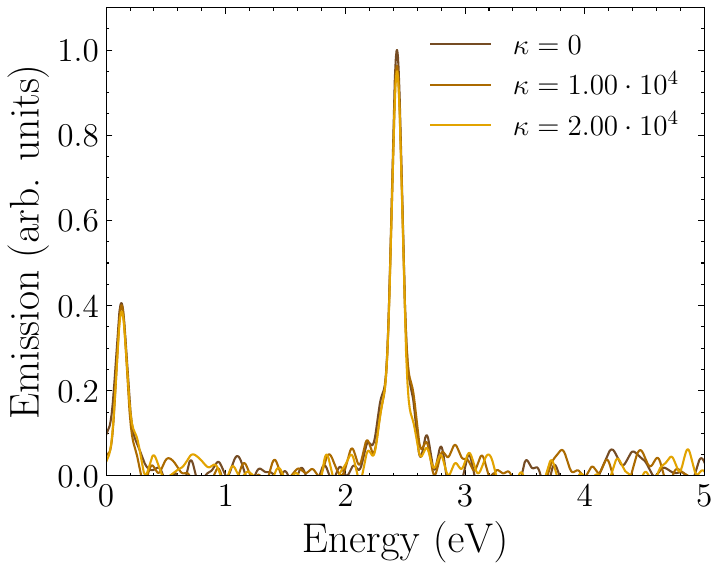}};
        \node[anchor=north west, xshift=11mm, yshift=-3mm, font=\large] at (img1.north west) {a)};
    \end{tikzpicture}
    
    \begin{tikzpicture}
        \node (img2) {\includegraphics[width=0.7\columnwidth]{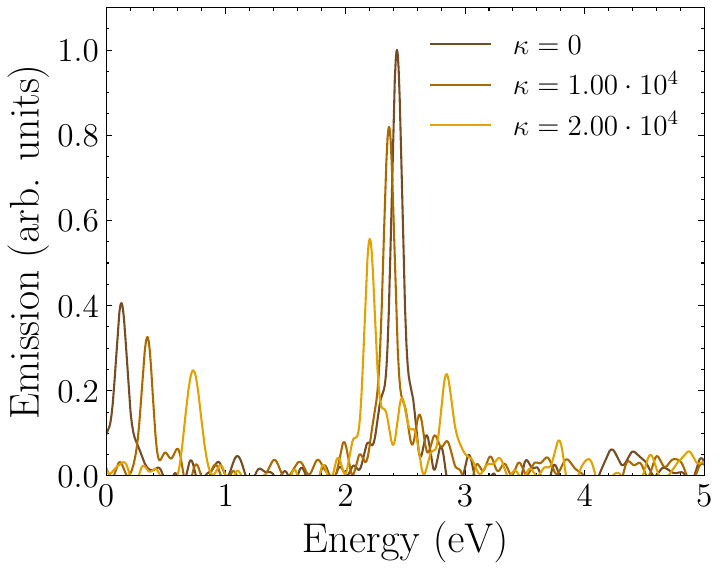}};
        \node[anchor=north west, xshift=11mm, yshift=-3mm, font=\large] at (img2.north west) {b)};
    \end{tikzpicture}

    \caption{Spectroscopy algorithm simulation for the 9-orbital boron vacancy defect, starting from the singlet state, and evolving under a) $H = H_\eff + \kappa H_\SOC^{1,0}$, and b) $H = H_\eff + \kappa (H_\SOC^{1,1} + H_\SOC^{1,-1})$. The simulation uses a second-order product formula, $S_{\text{Had}} = 3000$, $j_{\max} = 500$, $\eta = 0.002$, $\|H\|_w=1$ and $h = \tau = \pi / (2\|H\|_w)$. Note that the axial ISC rate is weak, as evidenced by the lack of modification of the spectral intensity in panel (a): meanwhile the non-axial ISC rate is appreciably larger, leading to noticeable spectral intensity leakage to the triplet spin sector, preferentially from the third singlet state.}
    \label{fig:spectroscopy_example}
\end{figure}

Next, we test the performance of the algorithms directed at predicting ISC imbalance, starting with the spectroscopy-based approach. In \cref{fig:spectroscopy_example} we show simulated spectra for the boron vacancy computed using $H = H_\eff + \kappa H_\SOC^{1, 0}$ to target the axial ISC rate, and $H = H_\eff + \kappa (H_\SOC^{1, 1} + H_\SOC^{1, -1})$ to target the non-axial one, all within in the same $N=9$ spatial orbital active space.
Here a first-order product formula was used with parameters $S_{\text{Had}} = 3000$, $j_{\max} = 500$, $\eta = 2\cdot 10^{-3}$ Ha, $\|H\|_w=1$ Ha, and $h = \tau = \pi / (2\|H\|_w)$. The resulting spectra (\cref{fig:spectroscopy_example}) show that in the case of the non-axial rate (panel (b)), the peak intensity is reduced relative to the reference spectrum when the (boosted) SOC operator component is introduced. This reduction, proportional to the exact ISC rate, signals robust non-axial ISC occurring in the system, specifically from the third singlet state.
In contrast, panel (a) of the same figure shows no adjustment of peak intensities in the axial channel when the SOC operator is boosted by the same amount, leading to the conclusion that axial ISC is weak or at least much weaker than the non-axial one. These observations of the ISC rate imbalance---confirming ODMR activity in this spin defect---align with the classically computed results in \cref{table:me_soc}, thus validating the spectroscopy-based quantum algorithm.

Importantly, spectral changes initiated from the singlet spin sector serve as direct confirmation of ODMR-activity. 
However, if the initial state were a triplet, e.g., $S=1$, $M =1$, a reduction in peak intensity alone would not unambiguously indicate ISC. 
This is because the spin-tensor operator $H_\SOC^{1,\pm 1}$ also enables transitions within the triplet manifold, such as $S=1$, $M=1$ to $S=1$, $M=0$, which would produce qualitatively similar spectral changes.
To resolve the nature of the transitions, 
a $S_z$ shift to the Hamiltonian can be added, that is physically equivalent to introducing an external field $B$.
Such a shift would affect the denominator of the perturbative term in~\cref{eq:perturbed_eigenstates_HSOC,eq:perturbed_eigenstates_HSOC:degenerate}, and would allow us to recognize which eigenstates were involved in the intersystem crossing. 

The energy shifts observed in~\cref{fig:spectroscopy_example} impose a constraint on eigenstate identification.
The perturbation must not mix states excessively, and their energy displacement should remain a fraction of the relevant energy gaps. Achieving this in other defects may require higher algorithm precision with a lower Trotter error, larger $j_{\max}$, lower $\eta$ and $\tau$, and more shots.

\begin{figure}
    \centering
    \includegraphics[width=\linewidth]{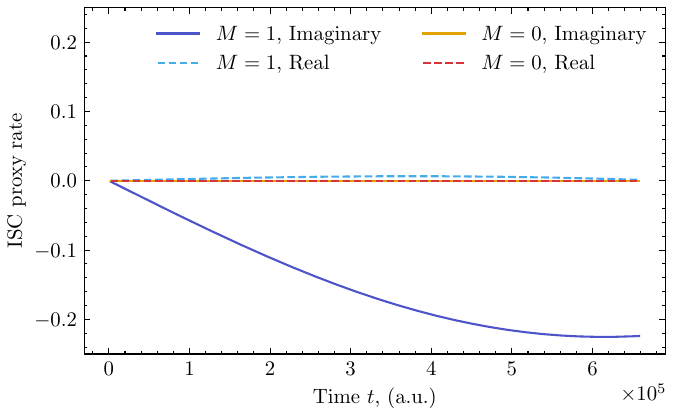}
    \caption{Simulation of the evolution-proxy algorithm for the $\text{V}_\text{B}^-$ spin defect using an active space of $N = 9$ spatial orbitals. The non-axial ISC amplitude proxy (blue) is much larger than the axial one (red) in this short-time simulation, the regime where the proxies are tightly proportional to the exact ISC rates, allowing us to conclude that there is a strong ISC rate imbalance and thus likely ODMR activity.}
    \label{fig:evolution_proxy}
\end{figure}

Finally, we simulate the evolution-proxy algorithm to probe ISC rate imbalance, again using the effective Hamiltonian with $N = 9$ spatial orbitals, following the procedure outlined in \cref{alg:evolution_proxy}. Three initial states $\ket{\psi_{S=0, M=0}}$,  $\ket{\psi_{S=1, M=0}}$ and $\ket{\psi_{S=1, M=1}}$ are prepared combining Sum of Slaters with QSP projection. These states are used to measure the corresponding matrix elements $\tilde k^\perp(t)$ and $\tilde k^z(t)$ in~\cref{eq:proxy_rates_nc_a,eq:proxy_rates_nc_b} which characterize spin-orbit-mediated ISC transitions and proportional to the exact ISC rates. The resulting proxy rates are shown in \cref{fig:evolution_proxy} in the short-time regime.
The significant imbalance between the axial and non-axial proxy rates at short times is indicative of a strong ISC rate imbalance, once again confirming the ODMR activity of the simulated defect in agreeing with the classical reference.

\subsection{Resource estimates~\label{ssec:resoureces_boron_vacancy}}

Using the approaches presented in \cref{sec:resource_estimation}, we now present concrete resource estimates for computing the optical response and ODMR activity of the boron vacancy in hexagonal boron nitride. With the Hamiltonian in hand, the main uncertainty in using the resource estimation formulas is the determination of the free parameters, which are the number of Hamiltonian CDF fragments $L$ (which we already fixed $L = N$, as argued in the previous section), simulation time step $\tau$, Trotter step $h$, the shots budget $S_{\text{Had}}$, and additionally for the spectroscopy-based algorithm only also the maximal evolution time $\tau j_{\max}$, broadening $\eta$, and the effective dipole support window $\|H\|_\omega$. For both algorithms, we will assume that we are preparing an initial state represented by a sum-of-Slaters with $D=10^4$ determinants~\cite{fomichev2024initial}.

While analytical error bounds presented in \cref{sec:resource_estimation} provide a valuable theoretical framework and allow to determine these parameters, we found the resulting bounds too loose in practice. To obtain tighter estimates, we conducted simulations with small active spaces, such as those presented in the previous section.

In both cases, we have already chosen the second-order Trotter product formula, and need to determine the Trotter step $h$ to be used. Applying the perturbative approach described in \cref{ssec:comm_based_algo_def}, in \cref{fig:trotter_error} of \cref{app:empirics} we plot the estimated perturbative error in the eigenstates for an active space of increasing size. The low overall magnitude of the observed Trotter error further justifies the choice of the (relatively low order) second order product formula for our resource estimates. Conservatively choosing $\epsilon_{\text{HS}} \leq 10^{-2}$ Ha and using the relation $h^{-1} \leq \sqrt{\epsilon_{\text{HS}}/\eta}$, we fix the Trotter step to keep the error below the desired threshold. The qubit cost indicated is the sum of the auxiliary qubits needed for the sum-of-Slaters technique~\cite{fomichev2024initial}, the auxiliary qubit in the Hadamard test, and the state register. For the parameters specific to each algorithm:

\textbf{Spectroscopy-based:} For this application, our simulations demonstrated that selecting $j_{\max} = 500$, $S_{\text{Had}} = 3000$, $\eta = 2$ mHa, $\|H\|_\omega = 1$ Ha, and $\tau = \pi / (2\|H\|_w)$ achieves the necessary spectral resolution. 

\textbf{Evolution-proxy:} If we target a precision of $\epsilon = 0.1$ in the modified Hadamard test, and assume a success probability upon projection of $\gamma^2 = |\braket{\psi_{\bm{D}, S, M}|\psi_{0,S,M}}|^2  = 0.7$ for each state $\ket{\psi_{0,S,M}}$, we need $\frac{4S_{\text{Had}}(\epsilon)}{\gamma^2} = \frac{16}{\epsilon^2 \gamma^2}\left(\frac{\gamma^2}{\alpha\beta}\right)^2\approx 2286$ circuit shots. 

Our resource estimates in~\cref{tab:resources_algorithms} assume that the energy windows of interest make $\lambda = 5$ Ha, and that the transition width $\Delta = 0.05$ Ha, so their ratio is approximatley $0.01$, which together with $\epsilon_{\text{QSP}} =0.01\ll \epsilon$ imply $d = 691$, computed with \texttt{pyqsp}~\cite{mrtc_unification_21,cdghs_finding_qsp_angles_20, haah_decomposition_19, gslw_qsvt_19, dmwl_efficient_phases_21}, see~\cref{fig:Heaviside}. An estimate of the Trotter error for small system sizes is computed in~\cref{fig:trotter_error}, which implies that taking $h = 1$ is sufficient to make $h^2\epsilon_{\text{HS}}\lesssim \Delta$.  

With this all the parameters needed to evaluate the cost of both the spectroscopy-based and evolution-proxy algorithms for the ISC rate imbalance have been specified. The resource estimates for a range of different sizes of active spaces are shown in \cref{tab:resources_algorithms}. The cost of the optical response algorithm is very nearly the cost of the spectroscopy-based algorithm for ISC, so we do not report it separately.

To contextualize our resource estimates, we compare them with previous studies. For instance, Baker {\it et al.}~\cite{baker2024simulating} estimated the cost of evaluating only the largest matrix element, $D_{i,0} = \braket{E_i|\bm{D}|E_0}$, using a qubitization-based variation of the QPE-sampling spectroscopy algorithm~\cite{fomichev2024simulating}. For a system with $N = 18$, they reported a cost of $1.1\cdot 10^{9}$ non-Clifford gates and 1633 logical qubits to obtain the largest dipole transition rate alone (not the full spectrum, as proposed here), without employing amplitude estimation. In contrast, the spectroscopy approach presented here reduces the logical qubit requirements by an order of magnitude while keeping the number of logical gates comparable. While prior constant-factor estimates on ISC rate calculations are not available, we note that the estimates we obtained here are comparable to the cost of estimating the optical response of the defect, with both being well-aligned with the range of expected capabilities of the early fault-tolerant hardware. 

\section{Conclusions}
\label{sec:conclusions}

Discovering novel material platforms that exhibit ODMR contrast is key to building the next generation of high-sensitivity magnetic field sensors. This paper presents two quantum algorithms designed to screen defect candidates for ODMR activity by detecting an imbalance in the ISC rates between excited states driven by spin-orbit coupling. 
The evolution-proxy algorithm achieves this by obtaining and comparing proxy quantities for axial and non-axial ISC rates. These proxies are defined as matrix elements of the time-evolution operator (under specific SOC components) between relevant initial and final states. This algorithm exploits the modified Hadamard test and efficient state preparation procedures to deduce the presence of an ISC rate imbalance. For the boron vacancy in hBN effective Hamiltonian with $N = 18$ spatial orbitals, it accomplishes this with 105 logical qubits and $4.41 \times 10^8$ Toffoli gates.
The spectroscopy-based method is more resource-demanding, each circuit needing the same number of qubits, but $2.21 \times 10^9$ Toffoli gates for the same model system. However, it provides additional information about the specific spin sublevels involved in the triplet-to-singlet intersystem crossing. This approach detects an ISC rate imbalance by looking for reductions in the optical spectrum intensity (relative to a reference spectrum) that are caused by ISC, then comparing the magnitude of the reductions for axial and non-axial SOC components. The spectra are obtained by time evolving the system under the Hamiltonian with spin-orbit coupling, whose strength is boosted to shorten the evolution time required to observe ISC effects.

Crucially, both algorithms leverage product formulas, achieving a significant reduction in qubit requirements---an order of magnitude improvement---compared to prior approaches. At the same time, the number of required non-Clifford gates remains well-aligned with the expected capabilities of early fault-tolerant hardware. While this advancement simplifies implementation, it introduces challenges in resource estimation, primarily due to the complexities of Trotter error. Future research should prioritize developing scalable methods that can produce tight estimates even for large system sizes, especially when the desired precision of the quantum simulation is high.

As new generations of sensors push beyond today's performance limits by relying more explicitly on quantum properties of matter,
accurately simulating quantum systems becomes ever more important. The current work is an early step towards not only using quantum computing to identify promising new materials for such devices, but also to begin characterizing their expected performance by directly modeling their response to external perturbation. As such devices continue to be developed and commercialized---for example for applications in GPS-free navigation, mineral prospecting, medical imaging, and beyond---computational resource demands for simulation will grow.
Classical computing faces challenges in modeling the quantum phenomena of strong correlation and entanglement essential to these sensors. Quantum computing is thus poised to become the main way to accurately model factors affecting device performance, including spin-orbit coupling and interactions with phonons. This will be key to improve prototype development, calibration, and overall performance.

\section{Acknowledgments}

This research used resources of the National Energy Research Scientific Computing Center (NERSC), a Department of Energy User Facility using NERSC award DDR QIS-ERCAP ERCAP0032729.

\section{Data availability}

The data for this manuscript might be found in~\cite{article_data}.

\bibliography{main}

\clearpage
\newpage
\appendix

\crefalias{section}{appendix}
\onecolumngrid

\section{Empirical results}
\label{app:empirics}

\begin{figure}[H]
    \centering
    \includegraphics[width=0.5\linewidth]{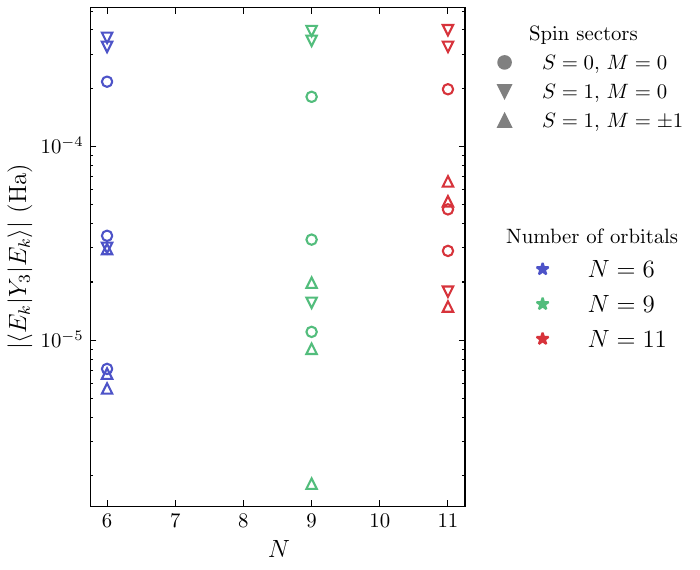}
    \caption{Computed Trotter error for a second order product formula for the CDF-BLISS electronic Hamiltonian. Each point represents a low-lying excited eigenstate, classified according to the spin sector it belongs to and the size of the active space.}
    \label{fig:trotter_error}
\end{figure}

\begin{figure*}[ht]
    \centering
\begin{minipage}{0.47\textwidth}
    \includegraphics[width=\linewidth]{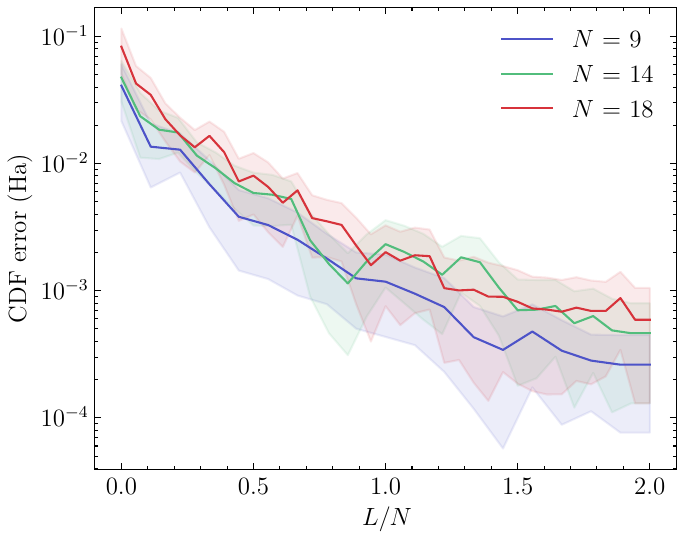}
    \caption{Mean and standard deviation error from approximating the electronic Hamiltonian with a CDF rank-factorized Hamiltonian, over the first 50 eigenvalues of the Hamiltonian. The x axis contains the number of fragments considered in the rank factorization, divided by the number of spatial orbitals. This plot empirically demonstrates that we have to scale $L = \tilde{O}(N)$ to keep the error constant~\cite{motta2021low,von2021quantum,lee2021even}.}
    \label{fig:cdf_error}
\end{minipage}
\hfill
\begin{minipage}{0.47\textwidth}
    \centering
    \includegraphics[width=\linewidth]{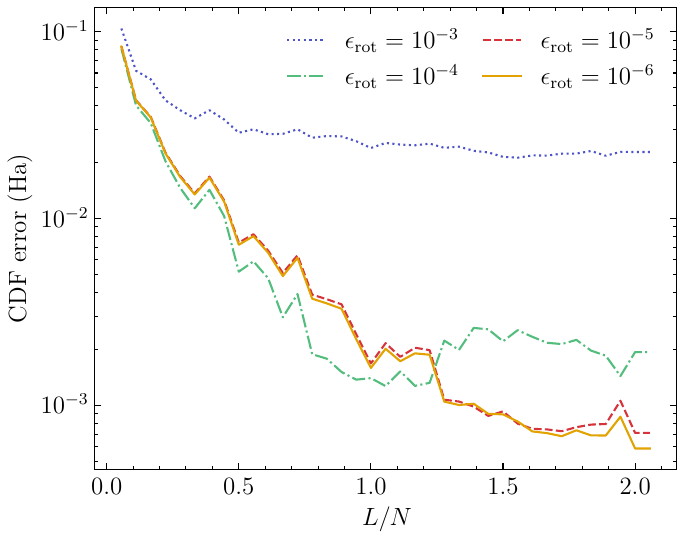}
    \caption{Mean error over the first 50 eigenvalues due to the truncation of the precision of the CDF approximation of the Hamiltonian, for N = 18 spatial orbitals. The x axis contains the number of fragments considered in the rank factorization, divided by the number of spatial orbitals. Setting the precision of each rotation at $\epsilon_{\text{rot}} = 10^{-4}$ provides close to milliHartree precision in the eigenvalues.}
    \label{fig:rotation_error}
\end{minipage}
\end{figure*}

\section{Matrix elements of the spin-orbit coupling operator}
\label{appx:soc_melements}
The matrix elements of the spin-orbit coupling operator defined in~\cref{eq:h_soc_op} are computed as
\begin{equation}
h_{pq}^\text{soc} = \frac{\alpha^2}{2} \sum_\sigma \int d\bm{r}~\phi_p^*(\bm{r}) \chi_{m_p}^*(\sigma) \times \left[ \sum_{I=1}^{N_A} \frac{Z_I}{|\bm{r}-\bm{R}_I|^3} [(\bm{r}-\bm{R}_I) \times \bm{p}] \cdot \bm{s} \right]~\phi_q(\bm{r}) \chi_{m_q}(\sigma),
\label{eq:h_soc_me_app}
\end{equation}
where $N_\text{A}$ is the number of atoms in the supercell. $Z_I$ and $\bm{R}_I$ are respectively the atomic number and coordinates of the $I$th atom, and $\bm{p}$ and $\bm{s}$ are the momentum and spin electron operators. In~\cref{eq:h_soc_me_app} $\chi_{m_p}(\sigma)$ is the spin function of the $p$th KS orbital $\phi_p(\bm{r})$. 

Making the change of variables $\bm{r} \rightarrow \bm{r}+\bm{R}_I$ we have
\begin{align}
h_{pq}^\text{soc} = \frac{\alpha^2}{2}\sum_\sigma \int d\bm{r}~\phi_p^*(\bm{r}+\bm{R}_I) \chi_{m_p}^*(\sigma) \left[ \sum_{I=1}^{N_A} \frac{Z_I}{r^3} (\bm{r} \times \bm{p}) \cdot \bm{s} \right]~\phi_q(\bm{r}+\bm{R}_I) \chi_{m_q}(\sigma),
\label{eq:1b_ls}
\end{align}
where $\bm{l}=\bm{r} \times \bm{p}$ is the angular momentum operator. Now, we express the $l_x$ and $l_y$ in terms of the ladder operators
\begin{align}
& l_x = \frac{1}{2}(l_+ + l_-), \label{eq:ladder1} \\
& l_y = \frac{1}{2i}(l_+ - l_-).
\label{eq:ladder2}
\end{align}
Eqs.~(\ref{eq:ladder1}-\ref{eq:ladder2}), which also apply to spin operator $s_x$ and $s_y$ allows us to write the product $\bm{l} \cdot \bm{s}$ as
\begin{equation}
\bm{l} \cdot \bm{s} = \frac{1}{2} (l_+ s_- + l_- s_+) + l_zs_z.
\label{eq:ls}
\end{equation}
By inserting \cref{eq:ls} into \cref{eq:1b_ls} we obtain
\begin{equation}
h_{pq}^\text{soc} = h_{pq}^{\text{soc}, \perp (1)} + h_{pq}^{\text{soc}, \perp (2)} + h_{pq}^{\text{soc}, z},
\label{eq:split_h_soc}
\end{equation}
where
\begin{align}
& h_{pq}^{\text{soc}, \perp (1)} = \frac{\alpha^2}{4} \sum_{I=1}^{N_A} \sum_\sigma \int d\bm{r}~\phi_p^*(\bm{r}+\bm{R}_I) \chi_{m_p}^*(\sigma) \left[ \frac{Z_I}{r^3}l_+ s_- \right] \phi_q(\bm{r}+\bm{R}_I) \chi_{m_q}(\sigma) \label{eq:soc_perp_1} \\
& h_{pq}^{\text{soc}, \perp (2)} = \frac{\alpha^2}{4} \sum_{I=1}^{N_A} \sum_\sigma \int d\bm{r}~\phi_p^*(\bm{r}+\bm{R}_I) \chi_{m_p}^*(\sigma) \left[ \frac{Z_I}{r^3} l_- s_+ \right] \phi_q(\bm{r}+\bm{R}_I) \chi_{m_q}(\sigma) \label{eq:soc_perp_2} \\
& h_{pq}^{\text{soc}, z} = \frac{\alpha^2}{2} \sum_{I=1}^{N_A} \sum_\sigma \int d\bm{r}~\phi_p^*(\bm{r}+\bm{R}_I) \chi_{m_p}^*(\sigma) \left[ \frac{Z_I}{r^3} ~ l_z s_z \right] \phi_q(\bm{r}+\bm{R}_K) \chi_{m_q}(\sigma).
\label{eq:soc_z}
\end{align}
Now, by separating the radial and spin integrals we obtain
\begin{equation}
h_{pq}^{\text{soc}, \perp (1)} = \frac{\alpha^2}{4} R_{pq}^{\perp (1)}~S_{m_p m_q}^{\perp (1)},~  h_{pq}^{\text{soc}, \perp (2)} = \frac{\alpha^2}{4} R_{pq}^{\perp (2)}~S_{m_p m_q}^{\perp (2)}, ~ h_{pq}^{\text{soc}, z} = \frac{\alpha^2}{2} R_{pq}^{z}~S_{m_p m_q}^{z},
\label{eq:radial_spin_ints}
\end{equation}
where
\begin{align}
& \kern-7pt R_{pq}^{\perp (1)} = \sum_{I=1}^{N_A} Z_I \int d\bm{r}~\phi_p^*(\bm{r}+\bm{R}_I) \left[ \frac{l_+}{r^3} \right] \phi_q(\bm{r}+\bm{R}_I),~S_{m_p m_q}^{\perp (1)} = \sum_\sigma \chi_{m_p}^*(\sigma) s_- \chi_{m_q}(\sigma) \label{eq:RSperp_1} \\
& \kern-7pt R_{pq}^{\perp (2)} = \sum_{I=1}^{N_A} Z_I \int d\bm{r}~\phi_p^*(\bm{r}+\bm{R}_I) \left[ \frac{l_-}{r^3} \right] \phi_q(\bm{r}+\bm{R}_I),~S_{m_p m_q}^{\perp (2)} = \sum_\sigma \chi_{m_p}^*(\sigma) s_+ \chi_{m_q}(\sigma) \label{eq:RSperp_2} \\
& \kern-7pt R_{pq}^z = \sum_{I=1}^{N_A} Z_I \int d\bm{r}~\phi_p^*(\bm{r}+\bm{R}_I) \left[ \frac{l_z}{r^3} \right] \phi_q(\bm{r}+\bm{R}_I),~S_{m_p m_q}^z = \sum_\sigma \chi_{m_p}^*(\sigma) s_z \chi_{m_q}(\sigma) \label{eq:RSperp_z}.
\end{align}
The Kohn-sham defect orbitals used to represent the observable are represented in a basis of plane wave basis functions
\begin{equation}
\phi_p(\bm{r}) = \frac{1}{\sqrt{\Omega}} \sum_i C_{ip} e^{i \bm{G}_i \cdot \bm{r}},
\label{eq:pw_basis}
\end{equation}
where $\Omega$ is the volume of the unit cell and $G_i$ denotes the $i$th reciprocal lattice vector. From~\cref{eq:pw_basis} we note that the spatial integrals entering $R_{pq}$ are computed as
\begin{equation}
\int d\bm{r}~\phi_p^*(\bm{r}+\bm{R}_I) \left[ \frac{O}{r^3} \right] \phi_q(\bm{r}+\bm{R}_I) = \frac{1}{\Omega} \sum_{i,j} C_{ip}^* C_{jq}~e^{i \bm{G}_{ji} \cdot \bm{R}_I} I_{ij}[O],
\end{equation}
where 
\begin{equation}
I_{ij}[O] = \int d\bm{r}~e^{-i \bm{G}_{i} \cdot \bm{r}} \left[ \frac{O}{r^3} \right] e^{i \bm{G}_{j} \cdot \bm{r}},
\end{equation}
with $O=\{l_-, l_+, l_z \}$ and $\bm{G}_{ji} = \bm{G}_j - \bm{G}_i$. Thus, we have
\begin{align}
I_{ij}[l_+] & = I_{ij}[l_x] + i I_{ij}[l_y] = \frac{4 \pi}{G_{ji}^2} \left[i(\bm{G}_{ji} \times \bm{G}_j)_x -  (\bm{G}_{ji} \times \bm{G}_j)_y \right]  \label{eq:lplus} \\
I_{ij}[l_-] & = I_{ij}[l_x] - i I_{ij}[l_y] = \frac{4 \pi}{G_{ji}^2} \left[i(\bm{G}_{ji} \times \bm{G}_j)_x +  (\bm{G}_{ji} \times \bm{G}_j)_y \right] \label{eq:lminus} \\
I_{ij}[l_z] & = \frac{4 \pi}{G_{ji}^2} i(\bm{G}_{ji} \times \bm{G}_j)_z \label{eq:lz},
\end{align}
where we have used the integral
\begin{equation}
\int d\bm{r} e^{ -i \bm{G} \cdot \bm{r} }  \left[ \frac{u}{r^3} \right] = \frac{4 \pi i G_u}{G^2},~~ \text{with}~~ u=x, y, z.
\label{eq:gpt_integral}
\end{equation}
Using Eqs. (\ref{eq:lplus}-\ref{eq:lz}) we obtain the following expression for the spatial matrix elements:
\begin{align}
& R_{pq}^{\perp (1)} = \frac{4\pi}{\Omega} \sum_{I=1}^{N_A} Z_I \sum_{i,j} C_{ip}^* C_{jq} \frac{e^{i \bm{G}_{ji} \cdot \bm{R}_I}}{G_{ji}^2} \left[ i(\bm{G}_{ji} \times \bm{G}_j)_x - (\bm{G}_{ji} \times \bm{G}_j)_y \right] \label{eq:R1_ij_perp} \\
& R_{pq}^{\perp (2)} = \frac{4\pi}{\Omega} \sum_I Z_I \sum_{i,j} C_{ip}^* C_{jq} \frac{e^{i \bm{G}_{ji} \cdot \bm{R}_I}}{G_{ji}^2} \left[ i (\bm{G}_{ji} \times \bm{G}_j)_x + (\bm{G}_{ji} \times \bm{G}_j)_y \right] \label{eq:R2_ij_perp} \\
& R_{pq}^z = \frac{4 \pi}{\Omega} \sum_{I=1}^{N_A} Z_I \sum_{i,j} C_{ip}^* C_{jq} \frac{e^{i \bm{G}_{ji} \cdot \bm{R}_I}}{G_{ji}^2} i(\bm{G}_{ji} \times \bm{G}_j)_z. \label{eq:R_ij_z}
\end{align}
On the other hand, using that
\begin{align}
& s_- \chi_{m_p}(\sigma) = \sqrt{(1/2+m_p)(1/2-m_p+1)} \chi_{m_p-1}(\sigma), \\
& s_+ \chi_{m_p}(\sigma) = \sqrt{(1/2-m_p)(1/2+m_p+1)} \chi_{m_p+1}(\sigma),
\end{align}
we obtain for the spin matrix elements
\begin{align}
& S_{m_p m_q}^{\perp (1)} = \delta_{m_p, m_q-1} \sqrt{(1/2+m_q)(1/2-m_q+1)} \label{eq:S1_perp} \\
& S_{m_p m_q}^{\perp (2)} = \delta_{m_p, m_q+1} \sqrt{(1/2-m_q)(1/2+m_q+1)} \label{eq:S2_perp} \\
& S_{m_p m_q}^z = m_q \delta_{m_p, m_q}.
\label{eq:Sz}
\end{align}

\section{SWAP test details \label{appx:swap}}
The swap test directly measures the squared magnitude of the overlap between the initial and final states of the spin-flip transition. The probability of measuring $\ket{0}$ is
\begin{equation}
    p(0) = \frac{1}{2} \left( 1 + |\braket{ \psi_{S=1, M} | e^{-itH_{\SOC}^{1,  M}}|\psi_{S=0} }|^2 \right).
\end{equation}
The transition probabilities are then calculated as
\begin{equation}
    |k''(t)|^2 = 2p(0) - 1.
\end{equation}
Statistical errors are suppressed through $\mathcal{O}(\epsilon^{-2})$ repeated measurements for precision $\epsilon$, ensuring robust estimation of spin-flip dynamics. 

\begin{figure*}
    \centering
    \begin{minipage}{0.45\textwidth}
        \centering
        \includegraphics[width=\linewidth]{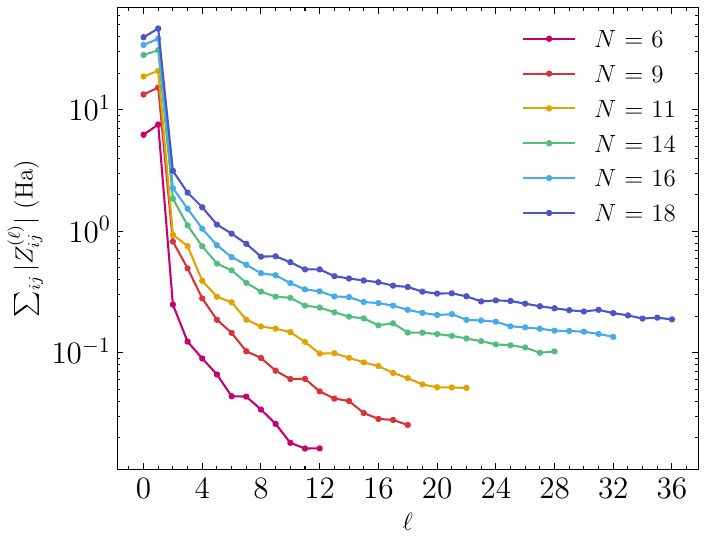}
        \caption{Size of the fragments of the Compressed Double Factorization approximation to $H_\eff$. The $\ell = 0$ term corresponds to the one-body term.}
        \label{fig:Zs}
    \end{minipage}
    \hfill
    \begin{minipage}{0.45\textwidth}
        \centering
        \[
            \Qcircuit @C=1em @R=.7em {
          \ket{0} &&&\qw & \gate{H} & \ctrl{9} & \ctrl{10} &\qw & \cdots & & \ctrl{13} &\gate{H} &\meter  \\
          \\
           &&&& & \qswap & \qw & \qw&\cdots &&\qw &  \qw& \qw\\
           &&&& & \qw& \qswap  & \qw&\cdots &&\qw &  \qw& \qw\\
         e^{-itH_{\SOC}^{1, M}} \ket{\psi_{S=0}}&&&&\vdots\\
          \\
             &&&&  & \qw & \qw&\qw & \cdots&& \qswap & \qw& \qw\\
          \\
          \\
        &&&\qw\qw& \qw & \qswap & \qw & \qw & \cdots &&\qw& \qw& \qw\\
           &&&\qw& \qw  & \qw& \qswap & \qw & \cdots &&\qw& \qw& \qw\\ 
         \ket{\psi_{1, M}}&&&&\vdots\\
           \\
             &&&\qw& \qw  & \qw & \qw&\qw & \cdots&& \qswap & \qw& \qw\\
            }
        \]
        \caption{Implementation of the swap test to compute the matrix element that defines the ISC rate. Since each state is prepared beforehand, the success probabilities do not enter the result, unlike the modified Hadamard test displayed in~\cref{fig:modified_Hadamard}.}
        \label{fig:swap_test}
    \end{minipage}
\end{figure*}

\section{Modified Hadamard test details \label{appx:modified_Hadamard}}
The modified Hadamard test provides an alternative method for measuring the real and imaginary components of the transition matrix element. This approach involves preparing a superposition state and applying controlled time evolution and measurements, as depicted in \cref{fig:modified_Hadamard}.

To elaborate, we begin with the following state:
\begin{align}\label{eq:aleph}
    \ket{\zeta} = \frac{\alpha}{\gamma\sqrt{2}}\ket{0}\ket{\phi} + \frac{\beta}{\gamma\sqrt{2}}\ket{1}\ket{\psi},
\end{align}
where $\gamma= \sqrt{\frac{|\alpha|^2+|\beta|^2}{2}}$.
Given an arbitrary Hermitian operator $\mathbf{O}$, we can measure
\begin{align}\label{eq:Hadamard_test_real}
     \braket{\zeta|X\otimes \mathbf{O}|\zeta} &= \frac{\alpha^*\beta}{2\gamma^2}\braket{\phi|\mathbf{O}|\psi} + \frac{\beta^*\alpha}{2\gamma^2}\braket{\psi|\mathbf{O}|\phi}
    =\text{Re}\left(\frac{\alpha^*\beta}{\gamma^2}\cdot\braket{\phi|\mathbf{O}|\psi}\right)
\end{align}
and
\begin{align}\label{eq:eq:Hadamard_test_imaginary}
     \braket{\zeta|Y\otimes \mathbf{O}|\zeta} &= i\frac{\alpha^*\beta}{2\gamma^2}\braket{\phi|\mathbf{O}|\psi}  -i\frac{\beta^*\alpha}{2\gamma^2}\braket{\psi|\mathbf{O}|\phi}
    =\text{Im}\left(\frac{\alpha^*\beta}{\gamma^2}\cdot\braket{\phi|\mathbf{O}|\psi}\right).
\end{align}

By setting $\mathbf{O} = \bm{1}$, $\ket{\psi} =e^{-i\frac{t}{2}H_{\SOC}^{1, M}}\ket{\psi_{S = 0}} $ and $\ket{\phi} = e^{+i\frac{t}{2}H_{\SOC}^{1, M}}\ket{\psi_{S = 1, M}}$, we can compute the matrix element up to constants $\alpha^*\beta$, see~\cref{fig:modified_Hadamard}.

The coefficients $\alpha$ and $\beta$ can be taken as real, and represent the amplitude of the state $\bm{D}\ket{E_{0}^{S, M}}$ in the target energy window:
\begin{align}\label{eq:alpha_AE}
    \ket{\psi_{\forall E, S=0}} &= \alpha \ket{\psi_{S=0}} + \sqrt{1-|\alpha|^2}\ket{\psi_{S=0}^\perp}\\
    \label{eq:beta_AE}
    \ket{\psi_{\forall E, S=1, M}} &= \beta \ket{\psi_{S=1, M}} + \sqrt{1-|\beta|^2}\ket{\psi_{S=1, M}^\perp}
\end{align}


The presence of $\alpha$ and $\beta$ in the Hadamard test result is not expected to pose a significant challenge. Although their values will differ depending on $M$, these variations are expected to be far smaller than the difference between matrix elements. Consequently, we should still reliably evaluate the existence of intersystem crossing. A substantial difference in $\alpha$ and $\beta$ would imply a small overlap between $\ket{\psi_{\forall E,S, M}}$ and the target energy window for some $M$, suggesting that the target eigenstate is not optically active. This, however, contradicts our initial assumptions.

Nevertheless, $\alpha$ and $\beta$ can be estimated at moderate additional gate cost. $\gamma^2/2$ is the success probability in the postselection step, see~\cref{fig:modified_Hadamard}. $\alpha$ can be independently estimated by sampling the success probability of the QSP state preparation of $\ket{\psi_{\forall E, S = 0}}$ in~\cref{eq:alpha_AE}. 
Finally, since $\alpha$, $\beta$ and $\gamma$ are real, $\beta$ can be derived from $\gamma^2 = \frac{\alpha^2 + \beta^2}{2}$. In this case, we have to rescale the precision of the Hadamard test by $\frac{\gamma^2}{\alpha \beta}$.

\section{Spectroscopy algorithm for degenerate eigenstates\label{appx:degenerate}}

Let us now assume that the triplet eigenstate is degenerate in the $M$ subspace.
We add an additional index to distinguish them $\ket{E_{l,j}}$. We orthogonalize the degenerate subspace:
\begin{align}\label{eq:orthogonalizing_degenerate_subspace}
\braket{E_{l,j_1}|H_{\SOC}^{1, M}|E_{l,j_2}} = \delta_{j_1,j_2}\braket{E_{l,j_1}|H_{\SOC}^{1, M}|E_{l,j_1}}
\end{align}
Then, assuming $H_{\SOC}^{1, M}$ lifts the degeneracy~\cite{Zwiebach18Perturbation},
\begin{align}\label{eq:perturbed_eigenstates_HSOC:degenerate}
    &\ket{E'_{l,j}} = \ket{E_{l,j}} - \kappa \sum_{k\neq l} \frac{\braket{E_k|H_{\SOC}^{1, M}|E_l}}{E_k-E_l}
    \ket{E_k}+ \kappa\sum_{g\neq j} \sum_{k\neq l}\frac{\braket{E_{l,g}|H_{\SOC}^{1, M}|E_{k}}\braket{E_{k}|H_{\SOC}^{1, M}|E_{l,j}}}{(E_k-E_l)(E_{l,j}^{(1)}-E_{l,g}^{(1)})}\ket{E_{l,g}}+O(\kappa^2)\\
    \label{eq:perturbed_eigenvalues_HSOC:degenerate}
   & E'_{l,j} = E_l  +\kappa\braket{E_{l,j}|H_{\SOC}^{1, M}|E_{l,j}}
    -\kappa^2 \sum_{k\neq l} \frac{|\braket{E_k|H_{\SOC}^{1, M}|E_{l,j}}|^2}{E_k-E_{l}}+ O(\kappa^3).
\end{align}
$E_{l,j}^{(1)}:=\braket{E_{l,j}|H_{\SOC}^{1, M}|E_{l,j}}$.
$\braket{E_{l,j}|H_{\SOC}^{1, M}|E_{l,j}}$ may vanish whenever $H_{\SOC}^{1, M}$ changes the total spin. However, due to~\cref{eq:orthogonalizing_degenerate_subspace}, $E_{l,j}$ are no longer eigenstates of $M$. As a result, $\braket{E_{0,S,M}|D_\rho|E^{(1)}_{l,j}}$ may not cancel in general. This does not preclude extracting intersystem crossing information across spin sectors. Note also that we may always artificially lift the degeneracy with an appropriately chosen shift by $S_z$.


\section{Implementing $\exp(-i t S^2)$\label{app:S^2}}

One important question is how to efficiently implement $\exp(-i t S^2)$. To do this we first observe that
\begin{equation}
    S^2 = S_x^2 +S_y^2 + S_z^2
\end{equation}
Fortunately, we know how to compute $S_z$,
\begin{align}
    S_z = \sum_k m_k \hat{n}_k
\end{align}
where $m_k$ is the projection number and $\hat{n}_k$ the occupancy. The square of this operator is two-body:
\begin{align}
    S_z^2 = \sum_{k,l} m_k m_l \hat{n}_k \hat{n}_l
\end{align}

Further, we also know that $S_x$ and $S_y$ are basis rotated versions of $S_z$. Thus, we can build a rank-3 CDF of the $S^2$ operator. An important question is what are the basis rotations that convert $S_x$ and $S_y$ into $S_z$. 

First we notice the definition of $S_x$ and $S_y$:
\begin{align}
    S_x\phi_{k,\uparrow} &= \phi_{k,\downarrow}\\
    S_x\phi_{k,\downarrow} &= \phi_{k,\uparrow}\\
    S_y\phi_{k,\uparrow} &= i \phi_{k,\downarrow}\\
    S_y\phi_{k,\downarrow} &= -i \phi_{k,\uparrow}\\
    S_z\phi_{k,\uparrow} &= \phi_{k,\uparrow}\\
    S_z\phi_{k,\downarrow} &= - \phi_{k,\downarrow}
\end{align}
As such, the basis rotation should be applied to each spatial orbital independently. Also,
\begin{align}
    S_x &= \begin{pmatrix}
        0 & 0 & 0 & 0\\
        0 & 0 & 1 & 0\\
        0 & 1 & 0 & 0\\
        0 & 0 & 0 & 0
    \end{pmatrix}\qquad
    S_y = \begin{pmatrix}
        0 & 0 & 0 & 0\\
        0 & 0 & i & 0\\
        0 & -i & 0 & 0\\
        0 & 0 & 0 & 0
    \end{pmatrix}\\
    S_z &= \begin{pmatrix}
        0 & 0 & 0 & 0\\
        0 & -1 & 0 & 0\\
        0 & 0 & 1 & 0\\
        0 & 0 & 0 & 0
    \end{pmatrix} = n_{\uparrow}- n_{\downarrow} = \frac{1-Z_{\uparrow}}{2} - \frac{1-Z_{\downarrow}}{2} \nonumber
\end{align}

We can find the basis that diagonalizes operators $S_x$ and $S_y$.
In particular these matrices diagonalize $S_x$ and $S_y$ respectively, because their columns are the eigenvectors of those operators:
\begin{align}
    U_x &= \frac{1}{\sqrt{2}}\begin{pmatrix}
        \sqrt{2} & 0 & 0& 0\\
         0 & 1 & 1 & 0\\
         0 & 1 & -1 & 0\\
         0 & 0 & 0& \sqrt{2}
    \end{pmatrix}\\
    U_y &=  \frac{1}{\sqrt{2}}\begin{pmatrix}
        \sqrt{2} & 0 & 0& 0\\
         0 & 1 & 1 & 0\\
         0 & -i & i & 0\\
         0 & 0 & 0& \sqrt{2}
    \end{pmatrix}
\end{align}
and
\begin{align}
    S_x = U_x E_x U_x^\dagger,\qquad
    S_y = U_y E_y U_y^\dagger.
\end{align}
where $E_x = E_y = -S_z$ are the eigenvalues. If we want to have exactly $S_z$ in the middle, without the minus sign, we can sandwich it with swap gates between the spin up and down register.

\begin{figure}[ht]
\begin{align}
\begin{array}{c}
    \Qcircuit @C=1em @R=.7em {
& \targ & \ctrl{1} & \ctrl{1} & \ctrl{1} & \targ& \qw \\
& \ctrl{-1} &\targ  & \gate{U} &\targ & \ctrl{-1}& \qw
}
\end{array}
\end{align}\caption{Circuit implementing the unitary $U'$ using a controlled version of the unitary $U$. When $U = X$ this circuit decomposes into the standard decomposition for a swap gate.} \label{fig:circuit_S2} 
\end{figure}
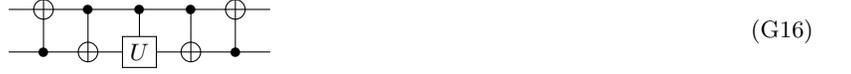

The question thus is how to implement these basis rotations. If we have a single qubit unitary 
\begin{align}
    U = \begin{pmatrix}
        a & b\\
        c & d
    \end{pmatrix}
\end{align}
we want to find a way to synthesize 
\begin{align}
    U' = \begin{pmatrix}
    1 & 0 & 0 & 0\\
    0 & a & b & 0\\
    0 & c & d & 0\\
    0 & 0 & 0 & 1\\
    \end{pmatrix}
\end{align}
The circuit in~\cref{fig:circuit_S2} does it.
The only question left is to find gates $U$ for the middle blocks in $U_x$ and $U_y$. Clearly, for $U_x$ it is the Hadamard gate, and it is also not hard to see that the $U_y$ is a Hadamard followed by an $S^\dagger$ gate. The necessary controlled Hadamard can be decomposed as
\begin{align}
    \text{C-H} = (\mathbf{1}\otimes R_y(\pi/4)) (\text{C-Z})(\mathbf{1}\otimes R_y(\pi/4))
\end{align}
Similarly, a C-S$^\dagger$ can implemented with the methods in Ref.~\cite{kim2018efficient}, with $2$ $T$  gates, one $T^\dagger$ gate and two CNots.

The last question is how to implement $\exp(itS_z^2)$. Since
\begin{align}
    S_z^2 &= \left(\sum_k n_{k,\uparrow}-n_{k,\downarrow}\right)^2\\
    &= \sum_{\sigma,\tau}(-1)^{\sigma+\tau}\sum_{k,l} n_{k,\sigma}n_{k,\tau}\\
    &= \sum_{\sigma,\tau}(-1)^{\sigma+\tau}\sum_{k,l} \frac{1-z_{k,\sigma}}{2}\frac{1-z_{k,\tau}}{2}\\
    &= \frac{1}{4}\sum_{\sigma,\tau}(-1)^{\sigma+\tau}\sum_{k,l} z_{k,\sigma}z_{k,\tau}-\frac{1}{4}\sum_{\sigma,\tau}(-1)^{\sigma+\tau}\left(\sum_{k} z_{k,\sigma} + \sum_{l} z_{l,\tau}\right)+\frac{1}{4}\sum_{\sigma,\tau}(-1)^{\sigma+\tau}\sum_{k,l} \bm{1}.
\end{align}
The final term, a global phase, adds up to 0 because of the term $\sum_{\sigma, \tau}(-1)^{\sigma + \tau}$, where the meaning is $(-1)^{\sigma + \tau} = 1$ if $\tau = \sigma$ and $(-1)$ otherwise. Similarly, the one-body term cancels, because for a given choice of $\sigma$, we have to sum over $(-1)^{\sigma+\tau}$. Thus, only the two-body term survives. However, when $(k,\sigma) = (l,\tau)$, the product $z_{k,\sigma}z_{l,\tau} = \bm{1}$, which creates an additional global phase. Thus,
\begin{align}
    S_z^2 &= \sum_{(k,\sigma)\neq (l,\tau)} (-1)^{\sigma+\tau}z_{k,\sigma}z_{l,\tau} + \sum_{(k,\sigma)} \bm{1}= \sum_{(k,\sigma)\neq (l,\tau)} (-1)^{\sigma+\tau}z_{k,\sigma}z_{l,\tau} + 2N \bm{1}.
\end{align}


\end{document}